\documentclass[fleqn,usenatbib]{mnras}
\usepackage[LGR,T1]{fontenc}
\usepackage[utf8]{inputenc}

\usepackage[greek,english.british]{babel} 
\usepackage[autostyle, autopunct, english=british]{csquotes}

\newcommand*{\mytitle}{Shaken and stirred: the effects of turbulence and rotation on disc and outflow formation during the collapse of magnetised molecular cloud cores\xspace}
\newcommand*{\myauthor}{B. T. Lewis and M. R. Bate\xspace}
\newcommand*{\myrunhead}{Shaken and stirred\xspace}
\newcommand*{\mykeywords}{accretion, accretion discs --- MHD --- radiative transfer --- stars: formation --- stars: winds, outflows --- turbulence\xspace}
\newcommand*{\mykeywordspdf}{accretion; accretion discs, MHD, radiative transfer, stars: formation, stars: winds; outflows, turbulence}

\hypersetup{pdfauthor={\myauthor}, pdftitle={\mytitle}, pdfkeywords={\mykeywordspdf}, pdfsubject={\myrunhead}}
\newcommand*{\mylbp}{\citet{2015MNRAS.451.4807L}\xspace}
\newcommand*{\mylb}{\citet{2017MNRAS.467.3324L}\xspace}

\usepackage{xspace}

\usepackage{newtxtext,newtxmath}

\usepackage[cal=rsfso,calscaled=.96,scr=rsfs,scrscaled=.96]{mathalfa}

\let\oldtextgreek\textgreek
\newcommand*{\greekfont}{artemisia}

\renewcommand{\textgreek}[1]{{\fontfamily{\greekfont}\selectfont\oldtextgreek{#1}}}

\newcommand*{\dvibackend}{pdftex}

\newcommand*{\diffcolor}{White}
\usepackage[natural,\dvibackend,hyperref,dvipsnames]{xcolor}
\usepackage{soul}
\sethlcolor{\diffcolor}

\newcommand*{\hleqn}[1]{\colorbox{\diffcolor}{\hspace{1em}#1\hspace{1em}}}

\usepackage[\dvibackend]{graphicx}

\DeclareGraphicsExtensions{.eps,.png,.pdf}

\usepackage{nicefrac}

\usepackage[sepfour,addmissingzero,noaddplus,autolanguage,np]{numprint}

\usepackage{cleveref}

\Crefname{figure}{Figure}{Figures}
\crefname{figure}{figure}{figures}
\Crefname{equation}{Equation}{Equations}
\crefname{equation}{equation}{equations}

\usepackage{aas_macros}
\usepackage{amsmath}
\usepackage{amssymb}

\usepackage{gensymb}
\usepackage{empheq}

\usepackage{setspace}

\title[\myrunhead{}]{\mytitle{}}
\author[\myauthor{}]{Benjamin T. Lewis\thanks{E--mail: \href{mailto:benjamin.lewis@rit.edu}{benjamin.lewis@rit.edu}} and  Matthew R. Bate\thanks{E--mail: \href{mailto:mbate@astro.ex.ac.uk}{mbate@astro.ex.ac.uk}}\\
School of Physics and Astronomy, University of Exeter, Stocker Road, Exeter EX4 4QL}

\input{macros}

\newcommand*{\philtrans}{Phil. Trans. R. Soc.}
\newcommand*{\revmodphys}{Rev. Mod. Phys.}
\newcommand*{\jcompphys}{J. Comput. Phys.}
\newcommand*{\computphyscommun}{Comput. Phys. Commun.}
\newcommand*{\computscieng}{Comput. Sci. Eng.}

\begin{document}

\date{In original form 2017 July 11.}

\pagerange{\pageref{firstpage}--\pageref{lastpage}} \pubyear{2017}

\maketitle

\label{firstpage}

\begin{abstract}
We present the results of eighteen magnetohydrodynamical calculations of the collapse of a molecular cloud core to form a protostar. Some calculations include radiative transfer in the flux limited diffusion approximation while others employ a barotropic equation of state. We cover a wide parameter space, with mass--to--flux ratios ranging from $\mu = 5$ to $20$; initial turbulent amplitudes ranging from a laminar calculation (i.e. where the Mach number, $\Mach = 0$) to transonic $\Mach = 1$; and initial rotation rates from $\beta_\rmn{rot} = 0.005$ to $0.02$.
We first show that using a radiative transfer scheme produces warmer pseudo–discs than the barotropic equation of state, making them more stable. We then \enquote{shake} the core by increasing the initial turbulent velocity field, and find that at all three mass-to-flux ratios transonic cores are weakly bound and do not produce pseudo--discs; $\Mach =  0.3$ cores produce very disrupted discs; and $\Mach =  0.1$ cores produce discs broadly comparable to a laminar core. In our previous paper we showed that a pseudo--disc coupled with sufficent magnetic field is necessary to form a bipolar outflow. Here we show that only weakly turbulent cores exhibit collimated jets. We finally take the $\Mach = 1.0$, $\mu = 5$ core and \enquote{stir} it by increasing the initial angular momentum, finding that once the degree of rotational energy exceeds the turbulent energy in the core the disc returns, with a corresponding (though slower), outflow. These conclusions place constraints on the initial mixtures of rotation and turbulence in molecular cloud cores which are conducive to the formation of bipolar outflows early in the star formation process.
\end{abstract}


\section{Introduction}
\label{sec:intro}

Chaotic, turbulent, molecular clouds are the birthplaces of stars \citep{2002ApJ...576..870P,2007ARA&A..45..565M}. 
These large clouds evolve and ultimately structures therein \citep{2010A&A...518L.102A,2010A&A...518L.103M,2010A&A...518L..92W,2009ApJ...700.1609M} succumb to the \citet{1902RSPTA.199....1J} instability and collapse --- creating the the molecular cloud cores that ultimately form protostars \citep{1977ApJ...214..488S,2001A&A...365..440M}. Many open questions in astronomy are related to how these cores evolve, for example, how are multiple star systems formed and the related question of how is the very large initial angular momentum of the core reduced to the stellar angular momenta observed in fully evolved systems \citep{2013ApJ...774...82L}. Molecular clouds are comprised of a magnetised astrophysical plasma \citep{2012ARA&A..50...29C,2015ARA&A..53..583H} which results in magnetic effects playing an important role in the collapse of the cloud cores. Ultimately, once these cores collapse to protostellar densities, jets and other outflows are produced. 
These outflows, and particularly collimated jets, transport angular momentum away from the protostellar core and can therefore explain why stars are not observed to be rotating close to their break--up speed.
Consequently, an understanding of how the initial conditions in the molecular core affect the generation of outflows is an important area of study.

The gas within a galaxy is invariably turbulent \citep{1969ApJ...158..123R}. Consequently the molecular gas within a cloud also is; as seen, for example, in the Horsehead nebula (\citealp{2003AJ....125.2108P}, \citealp{2005A&A...440..909H}), and also more generally, \citep{2004ApJ...615L..45H}. These turbulent conditions must then cascade down to the scales of molecular cloud cores. In the process, the magnitude of the turbulence will decay from being supersonic (in a molecular cloud) to transonic or subsonic in the core. Several numerical calculations, including \citet{1981MNRAS.194..809L}, \citet{2000ApJ...535..887K}, \citet{2004RvMP...76..125M}, \citet{2009MNRAS.392..590B}, and \citet{2012MNRAS.419.3115B}, have demonstrated that the turbulent motions within the cloud are an essential part of its evolution. These motions produce a filamentary structure and then localised overdensities which can collapse and form stars \citep{1993ApJ...419L..29E}.

~

In our previous works, \citet*{2015MNRAS.451.4807L} and \citet{2017MNRAS.467.3324L}, we explored how the gravitational collapse of magnetised molecular cloud cores is affected by changing the geometry and strength of the initial magnetic field. However, all of the calculations presented in those two works used cloud cores which, although rotating, had completely laminar velocity fields. These two papers continued a long series of work in this field including \citet{2002ApJ...575..306T}, \citet{2004ApJ...616..266M}, \citet{2006ApJ...641..949B}, \citet{2007MNRAS.377...77P}, \citet*{2008ApJ...676.1088M}, \citet{2010MNRAS.409L..39C}, \citet*{2014MNRAS.437...77B}, and \citet*{2017MNRAS.470.1626K}. Most recently, although not discussed in this paper, attention has turned to including non--ideal MHD effects (e.g. \citealp*{2006ApJ...647L.151M,2007ApJ...670.1198M}, \citealp{2008MNRAS.385.2269P}, \citealp*{2014MNRAS.444.1104W}, , \citealp{2015ApJ...810L..26T} and \citealp*{2015ApJ...801..117T}, \citealp*{2016MNRAS.457.1037W}).

In this work, we present the results of a series of calculations involving turbulent cores with the same turbulent power spectrum as used in \citet{2009MNRAS.392..590B} et seq. and thereby bridge the gap between simulations of whole molecular clouds and of fully formed protostellar objects. This choice is also consistent with \citet{2000ApJ...543..822B}, who proposed that the observed line--widths of molecular cloud cores could be caused by turbulent motion following a power law (\ie where the wavenumber of the turbulence, $k$, follows $P\left(k\right) \propto k^n$) with $n\in\left[-3,-4\right]$. 
We show that the initial velocity field of a molecular cloud core has as important an impact as the initial magnetic field strength --- both of which are seeded by the overall molecular cloud. In addition we show that the initial angular momentum of the cloud is closely linked to the nature of any protostellar outflows (along with the magnetic field strength and geometry, as we observed in \citealp{2015MNRAS.451.4807L}).

We present the results of an ensemble of eighteen smoothed particle radiation magnetohydrodynamical (SPRMHD) calculations of the collapse of a one solar mass molecular cloud core. Each core studied has a different set of initial conditions, ranging from being almost laminar to transonic cores, and covering different initial angular momenta. We also take the opportunity to compare the use of an approximate barotropic equation of state with a flux--limited diffusion (FLD) radiative transfer (RT) scheme. 

In \cref{sec:method} we set out our numerical SPRMHD scheme and in \cref{sec:initial} we discuss the initial conditions used for the ensemble of calculations. In \cref{sec:mhdcontrarmhd} we compare the results from calculations using the barotropic equation of state with the full FLD RT scheme. Following this, in \cref{sec:first,sec:shaken,sec:stir} we discuss the common early phase of the collapse, the effect of increasing turbulence and the effect of increasing angular momentum on the subsequent evolution of the core, respectively. We then compare these results both to observations and existing theoretical work in \cref{sec:discussion}. Finally, our conclusions are summarised in \cref{sec:conclusions}.

\section{Method}
\label{sec:method}

Turbulent magnetised plasmas can only be explored using a numerical scheme. Therefore, these calculations are evolved by solving the equations of (flux--limited) radiation ideal magnetohydrodynamics with self-gravity, which comprise
\begin{equation}\label{eqn:continuity}
\Dt\rho = -\rho \nabla^i v^i \comma
\end{equation}

\begin{empheq}{equation}\label{eqn:momentum} 
\Dt v^i = -\frac{1}{\rho}\nabla^i\left( 
p + \frac{1}{8\uppi}B^2
\right) \delta^{ij} + \frac{1}{4\uppi}\frac{1}{\rho}\nabla^i B^i B^j 
- \nabla^i \phi
+ \frac{\kappa F^i}{c} \comma
\end{empheq}

\begin{equation}\label{eqn:induction}
\Dt \frac{B^i}{\rho} = \left( 
\frac{B^j}{\rho}\nabla^j
\right) v^i \comma
\end{equation}

\begin{equation}\label{eqn:flux}
\Dt \frac{E}{\rho} =
- \frac{1}{\rho}\nabla^i F^i
- \frac{1}{\rho}\nabla^i v^j P^{ij}
+ 4\uppi\kappa B\left(T_\rmn{fl}\right)
- c\kappa E
\comma
\end{equation}

\begin{equation}\label{eqn:internalenergy}
\Dt u =
- \frac{p}{\rho} \nabla^i v^i 
- 4\uppi\kappa B\left(T_\rmn{fl}\right)
+ c\kappa E \comma
\end{equation}

\begin{equation}\label{eqn:gravity}
\Laplace \phi = 4\uppi G \rho \fstop
\end{equation}

In \cref{eqn:continuity,eqn:momentum,eqn:induction,eqn:flux,eqn:internalenergy,eqn:gravity} and the rest of this paper, 
\begin{equation}\label{eqn:convective}
\Dt \equiv \partial_t + v^i\nabla^i 
\end{equation}
denotes the convective derivative operator, with $\partial_t$ being the partial derivative with respect to time, $\Laplace$ is the Laplacian operator; $\rho$, $p$, $v^i$, $B^i$, $F^i$, $E$ and $\phi$ are the density, hydrodynamic fluid pressure, velocity, magnetic, radiative flux, radiation energy density and gravitational potential fields respectively; $u$ is the specific energy of the fluid; $P^{ij}$ is the radiation pressure tensor; $\kappa$ is the fluid opacity; {\npfourdigitnosep$c = \np{2.9979E10}~\cm\unitsep\second^{-1}$} is the speed of light in vacuo; and {\npfourdigitnosep$G = \np{6.6726E-8}~\dyne\unitsep\cm^2\unitsep\gram^{-2}$} is the gravitational constant. Repeated indicies imply summation. We assume that the fluid is always in local thermodynamic equilibrium. As a consequence we incorporate the emission of radiation by the fluid using the Planck function
\begin{equation}\label{eqn:planck}
B\left(T_\rmn{fl}\right) = \frac{\sigma_\mathrm{B}T^4_\mathrm{fl}}{\uppi}\comma
\end{equation} 
where {\npfourdigitnosep$\sigma_\mathrm{B} = \np{5.6705E-5}~\erg\unitsep\cm^{-2}\unitsep\second^{-1}\unitsep\kelvin^{-4}$} is the Stefan-Boltzmann constant and $T_\mathrm{fl}$ is the temperature of the fluid. An additional temperature --- the radiation temperature, $T_\mathrm{rad}$, is also defined as
\begin{equation}\label{eqn:trad}
E = \frac{4\sigma_\mathrm{B}T^4_\mathrm{rad}}{c} \fstop
\end{equation}

The large number of physical properties evolved require an array of different numerical approaches. We start by discretising these equations by using a smoothed particle radiation magnetohydrodynamics scheme, originally invented by \citet{1977AJ.....82.1013L,1977MNRAS.181..375G} and as set out for ideal MHD by \citet{2004MNRAS.348..123P,2005MNRAS.364..384P} and then extend this basic method as follows.

To ensure the efficient use of resolution and therefore computational resources we use a self--consistent variable smoothing length, $h$, scheme \citep{2004MNRAS.348..139P} where
\begin{equation}\label{eqn:smoothing}
h = \eta \left(\frac{m}{\rho} \right)^{\frac{1}{\nu}} \comma
\end{equation}
where $m$ is the mass of the SPH particle, $\eta = 1.2$ for the cubic B--spline kernel \citep{1985JCoPh..60..253M}, and $\nu = 3$ is the number of spatial dimensions. The typical number of neighbor particles is then $N_\mathrm{ngh} \sim 53$.

The MHD scheme given in \citet{2004MNRAS.348..123P,2005MNRAS.364..384P} is unstable in certain numerical situations \citep*[see][]{LBT2016}. We stabilise the momentum equation against the tensile pairing instability using the source term correction proposed by \citet*{2001ApJ...561...82B} with the parameter $\chi$ \citep[see the discussions in][]{2012JCoPh.231.7214T,LBT2016} fixed at $1$. The induction equation in SPMHD does not maintain the solenoidal constraint -- i.e. that $\nabla^i B^i = 0$. In previous work we used the constrained hyperbolic divergence cleaning proposed by \citet{2012JCoPh.231.7214T} based on the earlier grid-based method of \citet{2002JCoPh.175..645D}. In this work we adopt an improvement of this technique, detailed in \citet*{2016JCoPh.322..326T}, which is invariant to gradients in the magnetosonic wave velocity of the fluid. As before we couple a scalar field, $\psi$, to the induction equation such that
\begin{equation}\label{eqn:psi_ind}
\left.\Dt B^i \right|_{\nabla^i B^i} = - \nabla^i \psi \fstop
\end{equation}
However, we instead evolve $\nicefrac{\psi}{c_\mathrm{c}}$ as opposed to simply $\psi$ according to
\begin{equation}\label{eqn:psic}
\Dt \frac{\psi}{c_\mathrm{c}} = -c_\rmn{c}^2\nabla^iB^i - \frac{\psi}{\tau} - \frac{1}{2}\psi\left( \nabla^iv^i \right) \fstop
\end{equation} 
We set the cleaning wave speed, $c_\mathrm{c}$, to the fastest magnetosonic wave velocity, \ie
\begin{empheq}{equation}\label{eqn:magwavespeed}
c_\mathrm{c}^2 = c_\mathrm{s}^2 + c_\mathrm{a}^2 = c_\rmn{s}^2 + \frac{B^2}{4\uppi\rho} \comma
\end{empheq}
where $c_\rmn{s}$ and $c_\rmn{a}$ are the sound and Alfv\'en speeds of the fluid, and the damping timescale to
\begin{equation}
\tau = \frac{h}{\sigma_\mathrm{c}c_\mathrm{c}} \comma
\end{equation}
with $\sigma_\mathrm{c} = \nicefrac{4}{5}$ so that the cleaning wave is critically damped. These two additions effectively stabilise our MHD method for the duration of our calculations.

To prevent inter--particle penetration and to capture shocks in the fluid, we add an artificial viscosity term to the momentum equation, and to capture discontinuities in the magnetic field we add an artificial resistivity term to the induction equation (and therefore our method is actually quasi-ideal MHD). We use the Riemann solver based artificial viscosity term of \citet{1997JCoPh.136..298M} with a switch \citep{1997JCoPh.136...41M} to reduce dissipation when the viscosity term is unnecessary with the parameter $\alpha_\mathrm{AV} \in [0.1,~1.0]$. For the artificial resistivity, we use the switch proposed in \citet{2013MNRAS.436.2810T} with the parameter $\alpha_\mathrm{AR} \in [0.0,~1.0]$.

Self--gravitational forces are computed using a binary tree and we also use this tree to find the list of neighbour particles \citep{1990ApJ...348..647B,1988CoPhC..48...97B}. The gravitational potential itself is then softened using the ordinary SPH smoothing kernel, varying with smoothing length as in \cref{eqn:smoothing} \citep{2007MNRAS.374.1347P}.

\Cref{eqn:momentum,eqn:flux,eqn:internalenergy} contain radiation hydrodynamic (RHD) components where
\begin{equation}\label{eqn:fld}
F^i = - \frac{c\lambda\left(R\right)}{\kappa\rho} \nabla^i E \fstop
\end{equation}
This describes a fluid and radiation (i.e. two--temperature) FLD RT scheme. We evolve this RT scheme using the same implicit method as in \citet*{2005MNRAS.364.1367W}, \citet{2006MNRAS.367...32W}, \citet{2014MNRAS.437...77B}. We use the flux limiter \citep{1981ApJ...248..321L}
\begin{equation}\label{eqn:fluxlimiter}
\lambda\left(R\right) = \frac{2 + R}{6 + 3R + R^2} \comma
\end{equation}
where 
\begin{equation}\label{eqn:fld_R}
R = \left|\frac{\nabla^i E}{\kappa\rho E} \right| \comma
\end{equation}
with the opacity represented by $\kappa$. We obtain values of $\kappa$ from the opacity tables of \citet{1975ApJS...29..363A} and \citet*{1985Icar...64..471P} as detailed in \citet{2006MNRAS.367...32W}.
This allows the temperature evolution of the fluid to be calculated and evolved --- high density hot fluid regions will radiate and transfer energy to their cooler surroundings. The Eddington tensor, 
\begin{equation}\label{eqn:eddington}
f^{ij} = \frac{1}{2}\left[1 - f\left(E\right)\right] \delta^{ij}
+ \frac{1}{2}\left[3f\left(E\right) - 1\right] \hat{n}^i\hat{n}^j \comma
\end{equation}
can be used to calculate the radiation pressure from the radiation energy density 
\begin{equation}\label{eqn:PeqfE}
P^{ij} = f^{ij}E \fstop
\end{equation}
Finally, the Eddington factor function, $f\left(E\right)$, in \cref{eqn:eddington} and the flux limiter in \cref{eqn:fluxlimiter} are connected by the relation
\begin{equation}\label{eqn:edd_flux_connection}
f\left(E\right) = \lambda\left(R\right) + \lambda^2\left(R\right)R^2\left(E\right) \fstop
\end{equation}

Two equations of state are used in this paper. In most of the calculations presented in this paper, we use the radiative equation of state \citep[see][]{2006MNRAS.367...32W} given by 
\begin{equation}
P\left(\rho,T_\rmn{fl}\right) = \frac{\mathscr{R}}{\mu_\rmn{mol}}\rho T_\rmn{fl}\comma
\end{equation}
where the gas constant is given by $\mathscr{R} = 8.3145\times 10^{7}~\erg~\kelvin^{-1}~\mol^{-1}$, we use a mean molecular weight of $\mu_\rmn{mol} = 2.38$ and the fluid temperature $T_\rmn{fl} = \nicefrac{u}{C_\rmn{v}}$, where $C_\rmn{v}$ represents the isochoric specific heat capacity. The mean molecular weight and heat capacity used represent a mixture of hydrogen and helium and incorporate the effects of hydrogen dissociation and the ionization of hydrogen and helium, but does not include any contribution from metals \citep[see][]{1975ApJ...199..619B}.
In some calculations we use the barotropic equation of state given by \citep[\cf][]{2008ApJ...676.1088M}
\begin{empheq}[left={P\left(\rho\right) = c^{2}_{\rmn{s},0} \empheqlbrace}]{equation}
\begin{aligned}
& \rho{}          &&\rho{} \leq \rho_{\rmn{c,1}} \\
& \rho_{\rmn{c,1}} \left( \frac{\rho{}}{\rho_{\rmn{c,1}}} \right)^{\frac{7}{5}} & \hspace{-5pt}\rho_{\rmn{c,1}} <~ & \rho{} \leq \rho_{\rmn{c,2}} \\
& \rho_{\rmn{c,1}} \left( \frac{\rho_{\rmn{c,2}}}{\rho_{\rmn{c,1}}} \right)^{\frac{7}{5}} \left( \frac{\rho{}}{\rho_{\rmn{c,2}}} \right)^{\frac{11}{10}} && \rho{} > \rho_{\rmn{c,2}} \comma 
\label{eqn:barotrope}
\end{aligned}
\end{empheq}
where $c_{\rmn{s},0}$ represents the initial sound speed of the calculation and with critical densities of $\rho_\rmn{c,1} = 10^{-14}~\udens$ and $\rho_\rmn{c,2} = 10^{-10}~\udens$ to allow us to compare an approximate treatment with the full FLD RT scheme.
The physical consequences of the choice of critical densities were discussed in \mylb.
 
The entire SPRMHD detailed here is then temporally evolved using a second-order Runge--Kutta--Fehlberg integrator \citep[denoted RK2(4) in ][]{RK4NASATR}. Each SPH particle has an individual time-step so that particles in lower density or otherwise numerically simpler regions use fewer computational resources.

Sink particles \citep*{1995MNRAS.277..362B} are inserted when the density in the simulation exceeds either $\rho_\mathrm{crit} = 10^{-10}~\dens$ or $10^{-5}~\dens$, depending on the specific initial conditions. 
In an RT calculation the standard sink particle insertion tests are disabled (for instance sinks may be inserted with a non-negative $\nabla^i \dot{v}^i$). The sink particle insertion tests of \citet{1995MNRAS.277..362B} were designed to ensure that the region where $\rho \geq \rho_\rmn{crit}$ was undergoing gravitational collapse.  However, when wishing to insert a sink particle within a first hydrostatic core using an accretion radius that is smaller than the first core, the region being replaced is not collapsing (it is hydrostatic) and, in fact, is not self-bound (i.e. the external pressure matters).  Thus, the tests are disabled.  This can, however, cause spurious fragmentation if the critical density is too low.  For the calculations in this paper, choosing $\rho_\mathrm{crit} = 10^{-10}~\dens$ is sufficient to avoid spurious fragmentation for slowly rotating cores, but may produce spurious fragmentation in more rapidly rotating cores where the highest density regions are more extended. Therefore, for the calculations with an initial solid--body rotation rate of $\geq 3.54\times10^{-13}~\rad~\second^{-1}$, we delay the insertion of sinks until after the second collapse phase to prevent spurious fragmentation.  Since the slowly rotating cases only produce a single sink particle, raising the critical density in these calculations would have no significant effect other than to increase the computational time.

We set the accretion radius of the sink particles to $r_\mathrm{acc} = 1~\au$ (for both critical densities). Any SPH particles that fall within this distance of a sink particle are removed from the simulation and their masses are added to the sink if the usual checks for boundness detailed in \citet{1995MNRAS.277..362B} are passed. We impose a further inner boundary at $0.1~\au$ within which particles are accreted unconditionally to prevent small timesteps arising from large gravitational accelerations close to the sink particle. Our choice of accretion radius is a compromise between numerical efficiency --- a larger radius results in a shorter run--time --- and incorporating as much of the first hydrostatic core as possible (a first hydrostatic core has a typical radius of $\approx 4~\au$ \citep{1969MNRAS.145..271L}).  Previous work\citep[viz.][]{2012MNRAS.423L..45P,2015MNRAS.451.4807L} has shown that reducing the radius from $5~\au$ to $1~\au$ increases the jet velocity slightly from $\left|v_z\right| \approx 5~\kvelo$ to $8~\kvelo$, but has no other significant effect. 

Other than through gravity, the sink particles do not interact with other SPH particles: importantly they do not emit radiation (they are \enquote{cold}) and have no magnetic field components. Since the sink particle has no magnetic field, the field carried by any accreted particle is effectively destroyed, i.e. eliminated from the calculation rather than incorporated into the properties of the sink particle. The physical consequences of this infelicity have been discussed previously \citep[e.g. by][]{LBT2016}, and the primary effect on the calculation is to render a robust divergence cleaning scheme even more essential.

The calculations were performed using \textsc{sphng}, a three-dimensional hybrid \textsc{mpi} and \textsc{openmp} smoothed particle radiation magnetohydrodynamics code originating from \citet{X} and subsequently significantly modified by \citet{1995MNRAS.277..362B}, \citet{2006MNRAS.367...32W}, \citet{2007MNRAS.377...77P}, and others. Each individual calculation was completed on a pair of 6--core hyper--threaded CPUs (giving a total of 24 execution threads), taking between 16 and 32 days of wall time (approximately 553,000 and 1,100,000 core hours) depending on the initial conditions used.

\begin{table*}
\centering
\caption{Initial conditions for the calculations presented in this paper. $\mu$ is the mass-to-flux ratio (see \cref{eqn:mass2flux}, $B_0$ is the initial magnetic field strength, $\Mach$ is the initial turbulent Mach number, $\Omega_0$ is the initial angular speed of the core; $\beta_\rmn{mag}$, $\beta_\rmn{turb}$ and $\beta_\rmn{rot}$ are the ratios of magnetic, turbulent kinetic and rotational kinetic energy to gravitational energy (\cref{eqn:betamag,eqn:betaturb,eqn:betarot}), and $\beta_\rmn{tot}$ is the ratio of all the non--thermal supporting energy sources to the gravitational energy (\cref{eqn:betatot}). Note that calculations $\upmu$05--M10--r001(R) and $\upmu$05--M10--r002(R) use $\rho_\rmn{crit} = 10^{-5}\udens$, and all other calculations use $\rho_\rmn{crit} = 10^{-10}\udens$ as the critical density for sink particle creation. \hl{$\alpha_0 = 0.45$} (see \cref{eqn:alpha0}) throughout.}
\label{tbl:initialcond}
\begin{tabular}{l|lrrrrrrrrr}
Calculation Name  & Equation of State & $\mu$ & $B_0$ $\left[\mugauss\right]$ & $\beta_\rmn{mag}$ & $\Mach$ & $\beta_\rmn{turb}$ & $\Omega_0$ $\left[\times 10^{-13}\rads\right]$ & $\beta_\rmn{rot}$ & $\beta_\rmn{tot}$ & $\alpha_0 + \beta_\rmn{tot}$   \\ \hline
$\upmu$05-M00(R) & FLD RT       & 5  & 163 & 0.071  & 0   & 0      & 1.77 & 0.005 & 0.076  & \hl{0.526} \\
$\upmu$05-M01(M) & Barotropic   & 5  & 163 & 0.071  & 0.1 & 0.0013 & 1.77 & 0.005 & 0.077  & \hl{0.527} \\
$\upmu$05-M01(R) & FLD RT       & 5  & 163 & 0.071  & 0.1 & 0.0013 & 1.77 & 0.005 & 0.077  & \hl{0.527} \\
$\upmu$05-M03(M) & Barotropic   & 5  & 163 & 0.071  & 0.3 & 0.012  & 1.77 & 0.005 & 0.088  & \hl{0.538} \\
$\upmu$05-M03(R) & FLD RT       & 5  & 163 & 0.071  & 0.3 & 0.012  & 1.77 & 0.005 & 0.088  & \hl{0.538} \\
$\upmu$05-M10(M) & Barotropic   & 5  & 163 & 0.071  & 1.0 & 0.13   & 1.77 & 0.005 & 0.206  & \hl{0.656} \\
$\upmu$05-M10(R) & FLD RT       & 5  & 163 & 0.071  & 1.0 & 0.13   & 1.77 & 0.005 & 0.206  & \hl{0.656} \\
$\upmu$05-M10-r001(R) & FLD RT  & 5  & 163 & 0.071  & 1.0 & 0.13   & 3.54 & 0.01  & 0.211  & \hl{0.661} \\
$\upmu$05-M10-r002(R) & FLD RT  & 5  & 163 & 0.071  & 1.0 & 0.13   & 7.08 & 0.02  & 0.221  & \hl{0.671} \\
$\upmu$10-M00(R) & FLD RT       & 10 &  81 & 0.018  & 0   & 0      & 1.77 & 0.005 & 0.023  & \hl{0.473} \\
$\upmu$10-M01(R) & FLD RT       & 10 &  81 & 0.018  & 0.1 & 0.0013 & 1.77 & 0.005 & 0.024  & \hl{0.474} \\
$\upmu$10-M03(R) & FLD RT       & 10 &  81 & 0.018  & 0.3 & 0.012  & 1.77 & 0.005 & 0.035  & \hl{0.485} \\
$\upmu$10-M10(R) & FLD RT       & 10 &  81 & 0.018  & 1.0 & 0.13   & 1.77 & 0.005 & 0.153  & \hl{0.603} \\
$\upmu$20-M00(R) & FLD RT       & 20 &  41 & 0.0045 & 0   & 0      & 1.77 & 0.005 & 0.0095 & \hl{0.460} \\
$\upmu$20-M01(R) & FLD RT       & 20 &  41 & 0.0045 & 0.1 & 0.0013 & 1.77 & 0.005 & 0.0108 & \hl{0.461} \\
$\upmu$20-M03(R) & FLD RT       & 20 &  41 & 0.0045 & 0.3 & 0.012  & 1.77 & 0.005 & 0.0215 & \hl{0.472} \\
$\upmu$20-M10(M) & Barotropic   & 20 &  41 & 0.0045 & 1.0 & 0.13   & 1.77 & 0.005 & 0.140  & \hl{0.590} \\
$\upmu$20-M10(R) & FLD RT       & 20 &  41 & 0.0045 & 1.0 & 0.13   & 1.77 & 0.005 & 0.140  & \hl{0.590} \\
\end{tabular}
\end{table*}

\section{Initial Conditions}
\label{sec:initial}

We adopt initial conditions that are broadly similar to \mylb{}, which follow the general approach seen in \citet{2012MNRAS.423L..45P}, \citet*{2014MNRAS.437...77B}, \citet{2015MNRAS.451.4807L}   and others. However, in contrast to these works we add a turbulent velocity field and also vary the initial angular momentum of the molecular cloud core.

A cold uniform density sphere of SPH particles is placed within a warm low density container medium with the SPH particles initially placed on a cubic lattice, and the 
whole structure is then placed in a periodic box. This numerical set--up is similar to that used by \citet{HOSKING2002}; alternative arrangements which use, for example, 
Bonnor--Ebert spheres \citep{1956MNRAS.116..351B,1955ZA.....37..217E} instead of a uniform density distribution include \citet{2006ApJ...641..949B}. We use ca. 
$3\times 10^6$ SPH particles for the sphere (this is double the mass resolution in \mylbp and \mylb), which represents the molecular cloud core, and ca $1.5\times 10^6$ 
particles for the surrounding medium. The core has a radius of $r_\mathrm{core} = 4\times 10^{16}~\cm$ and a mass of $M_\mathrm{core} = 1~\solarm$, giving an initial 
uniform density of $\rho_{\mathrm{core},0} = 7.4\times 10^{-18}~\dens$. The initial sound speed of the core is set to $c_{\mathrm{s},0} = 2.2\times 10^4~\velo$ giving the core an initial 
fluid temperature of $T_{\mathrm{fl,core,}0} = 13.8~\kelvin$. The core and container medium are in pressure equilibrium so the low-density container medium has an initial 
fluid temperature of $\approx 400~\kelvin$.  When using radiative transfer, the radiation temperature, $T_{\rm rad}$, is initialised to 13.8~K in both regions.  Both the values of the fluid and radiation temperatures are held constant in the container medium throughout the calculations.

Our mass resolution significantly exceeds the resolution criteria in \citet{1997MNRAS.288.1060B} (which require a Jean's mass be resolved by at least $2N_\rmn{ngh}$ particles) but also ensures that we correctly capture magnetised turbulence since \citet{2011ApJ...731...62F} observed this requires significantly higher resolution than that required in hydrodynamic problems. 

The initial magnetic field is set uniformly throughout the core and container medium and is aligned to the $z$-axis (so $\vartheta = 0\degree$ and $B^{z}_0 = B_0$; \citealp{2015MNRAS.451.4807L}). The initial magnetic field strength is set according to the dimensionless mass--to--flux ratio, $\mu$, which is defined as
\begin{equation}\label{eqn:mass2flux}
\mu = \frac{\varpi_\mathrm{core}}{\varpi_\mathrm{crit}} \fstop
\end{equation}
The ratio between the magnetic pressure and self-gravitational forces, $\varpi_\mathrm{core}$, in a magnetised sphere is given by
\begin{equation}\label{eqn:mass2flux_numerator}
\varpi_\mathrm{core} = \frac{M_\mathrm{core}}{\uppi r^2_{\mathrm{core},0} B_0} \comma
\end{equation}
and the \enquote{critical} ratio at which these forces are in balance is given by

\begin{empheq}{equation}\label{eqn:mass2flux_denominator}
\varpi_\mathrm{crit} = \frac{c_1}{3\uppi}\sqrt{\frac{5}{G}} \approx 490~\gram~\gauss^{-1}~\cm^{-2}  \comma 
\end{empheq}
where we use $c_1 = 0.53$ as determined by \citet{1976ApJ...210..326M}. The field is periodic across the box, and continuous between the core and surrounding medium. 
This prevents numerical artifacts at the edge of the core even in very turbulent calculations.

The surrounding medium has zero initial velocity. The initial velocity profile of the core is produced by combining a turbulent component, $v^i_\mathrm{turb}$, and a solid-body rotation component, $v^i_\mathrm{rot}$, i.e.
\begin{equation}\label{eqn:initvel_summary}
v^i = v^i_\mathrm{turb} + v^i_\mathrm{rot} \fstop
\end{equation}
In contrast, \mylbp, \mylb, etc., only used a rotation component. To make comparisons with other calculations simpler, we use two dimensionless parameters, $\beta_\mathrm{turb}$ and $\beta_\mathrm{rot}$ representing the ratio of turbulent and rotational energy to gravitational potential energy respectively, to set the velocity field. These parameters are defined as
\begin{equation}\label{eqn:betarot}
\beta_\mathrm{rot} = \frac{1}{3}\frac{r_{\mathrm{core},0}^3 \Omega_0^2}{G M_{\mathrm{core},0}} \comma
\end{equation}
where $\Omega_0^2$ is the initial angular velocity of the core, and
\begin{equation}\label{eqn:betaturb}
\beta_\mathrm{turb} = \frac{1}{2} \frac{r_{\mathrm{core},0}\bar{v}^2_\rmn{turb}}{G M_{\mathrm{core},0}}\fstop
\end{equation}
$\bar{v}_\rmn{turb}$ is the root mean square (RMS) velocity of the turbulence, since the velocity of any individual particle will vary according to the Gaussian normal distribution discussed below, and is related to the Mach number, $\Mach$, of the turbulence via
\begin{equation}
\bar{v}_\rmn{turb} = \Mach c_\rmn{s} \comma
\end{equation}
where $c_\rmn{s}$ is the sound speed of the fluid (which is initially spatially constant). We also define the ratio of magnetic energy to gravitational potential energy as

\begin{empheq}{equation}\label{eqn:betamag}
\beta_\rmn{mag} = \frac{5}{18} \frac{B^2r_{\rm core,0}^4}{GM^2}
\end{empheq}
Using this we obtain values of $\beta_\rmn{mag} = 0.071, 0.018, \text{and~} 0.0045$ for $\mu = 5$, $10$, and $20$ respectively. We can then define the ratio of all the forms of non--thermal support --- magnetic pressure, turbulent kinetic, and rotational energy --- against the gravitational self--potential as
\begin{equation}\label{eqn:betatot}
\beta_\rmn{tot} = \beta_\rmn{mag} + \beta_\rmn{turb} + \beta_\rmn{rot}\fstop
\end{equation}
Using this we obtain values of $\beta_\rmn{tot}$ ranging from $0.0095$ for a $\mu = 20$, $\Mach = 0.1$ calculation to $0.221$ for $\mu = 5$, $\Mach = 1.0$, $\beta_\rmn{rot} = 0.02$ calculation, \ie ranging from a very super--critical core to one which less strongly bound. Our range of values of $\beta_\rmn{rot}$, from $0.005$ to $0.02$, is consistent with \citet{1993ApJ...406..528G} (see, in particular, the distribution in figure 11 therein), albeit with slightly less angular momentum than proposed by \citet{2000ApJ...543..822B}.

For completeness we can also calculate the ratio of thermal energy to the magnitude of the gravitational potential energy as
\begin{empheq}[box=\hleqn]{equation}\label{eqn:alpha0}
\alpha_0 = \frac{5}{2}\frac{\mathscr{R} T_\rmn{fl} r_{\rm core,0}}{\mu_\rmn{mol} G M}.
\end{empheq}
In all the calculations presented in this paper we have \hl{$\alpha_0 = 0.45$}. The sum, $\alpha_0 + \beta_\rmn{tot}$ ranges from \hl{$0.460$} to \hl{$0.671$}, i.e. from reasonably bound to a more weakly bound core. In a turbulent calculation it is likely that there will be regions, even in a core with $\alpha_0 + \beta_\rmn{tot} =~$\hl{$ 0.671$}, where locally the turbulent amplitude is sufficent to cause material to unbind. Conversely, although not covered in this paper, even if the sum $\alpha_0 + \beta_\rmn{tot} > 1$ there would likely be regions where the local turbulence is \textit{low} enough that these regions would remain bound.

\begin{figure}
\centering
\includegraphics[width=6.9cm]{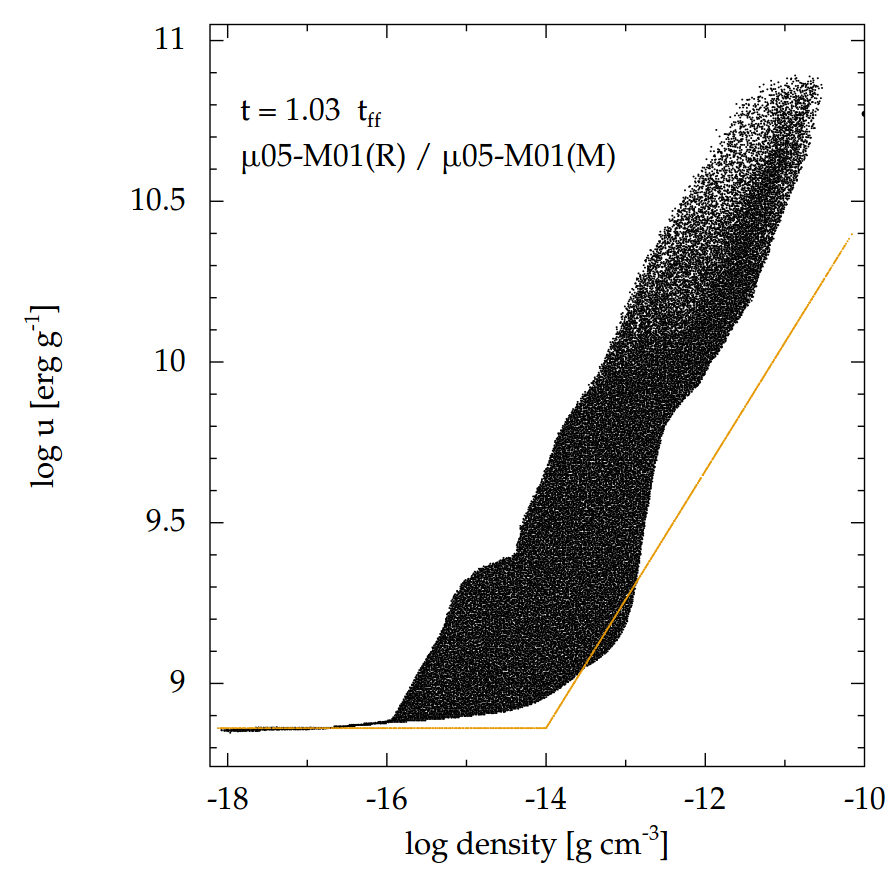}
\caption{Specific internal energy, $u$, as a function of the fluid density at $t = 1.03\thinspace\tff$ for all SPH particles within core of the $\upmu$05--M01(R) calculation (black dots) and $\upmu$05--M01(M) calculation (orange dots). The radiative transfer scheme causes a generally greater, and earlier, increase in $u$ compared to the barotropic equation of state. The latter is fixed until the first critical density ($\rho = \np{1E-14}\thinspace\udens$) is reached and linear thereafter; if regions of the fluid were to exceed the second critical density the slope of this line would change. The radiative transfer scheme is not a simple function of the density however. (see, \eg the \enquote{knee} in the temperature distribution between $\rho\approx\np{1E-13}\thinspace\udens$ and $\rho\approx\np{1E-12}\thinspace\udens$).
\label{fig:trho}}
\end{figure}

\begin{figure}
\centering{}
\includegraphics[width=0.5\textwidth]{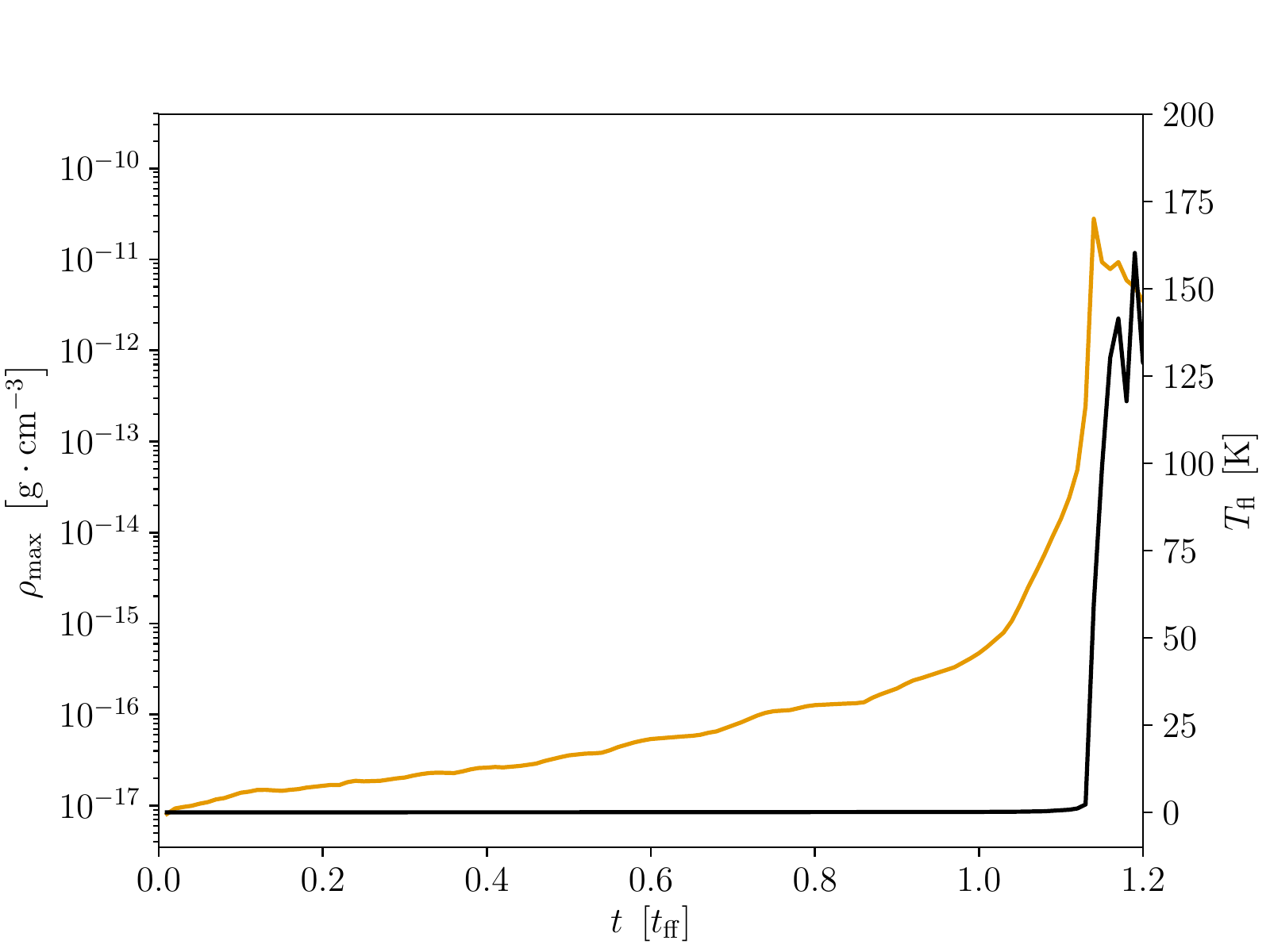}

\caption{\label{fig:tfluid-maxdens} Evolution of the maximum fluid density (solid orange line) and maximum fluid temperature (solid black line) for $\mu = 5$ with a $\Mach = 1.0$ initial turbulence field. In the early phase of the calculation, when the maximum density in the cloud core is $\rho \ll 10^{-12}~\udens$ the temperature of the core is effectively isothermal with $T = 13.8~\text{K}$. In this phase the core is transparent to the radiation energy produced from the gravitational collapse and consequently negligible heating occurs. At higher densities the fluid is heated and this energy is then transported by the radiative transfer scheme through the fluid.}
\end{figure}

The rotational component of the velocity field is provided by solid body rotation about the $z$--axis. The turbulent component is provided by giving every particle in the core an initial velocity at $t = 0$ --- the turbulence is not driven and therefore decays over time. We impose turbulence in a similar manner to \citet*{2001ApJ...546..980O} and \citet*{2003MNRAS.339..577B}. A uniform $128^3$ grid of velocities was generated and the initial particle velocities were interpolated from this, multiplied by a co--efficient to produce the correct overall turbulent Mach number. The generated grid represents a divergence--free (i.e. our turbulence is not compressive, cf. \citealp{2010A&A...512A..81F}) random Gaussian field with a power spectrum which follows $P\left(k\right) \propto k^{-4}$, where $k$ is the wavenumber, consistent with \citet{2000ApJ...543..822B}.  
The velocity dispersion, $\sigma$, of such a distribution follows scaling law of \citet{1981MNRAS.194..809L}, with $\sigma\left(r\right) \propto \sqrt{r}$, where $r$ is the distance.

We note that our initial conditions differ from \citet*{2017arXiv170309139M}, which use a Bonnor--Ebert sphere and apply \textit{only} an initially turbulent velocity field, rather than a superposition of a turbulent and a rotational field. 

This provides a wide range of potential initial conditions which can be explored. 
We present eighteen calculations in total, with the initial conditions as set out in \cref{tbl:initialcond}. These calculations span mass--to--flux ratios from $\mu = 5$ to $20$, turbulent velocities from $\Mach = 0$ (a laminar core) to $\Mach = 1$ (a transonic core), rotational velocities from $\beta_\rmn{rot} = 0.005$ to $0.02$, and include both barotropic and FLD RT formalisms. In this way, we sample a wide area of the parameter space, from a tightly bound, weakly magnetized, laminar core to weakly bound, transonic, strongly magnetized, and rapidly rotating core.

\begin{figure}
\centering{}
\includegraphics[width=0.44\textwidth]{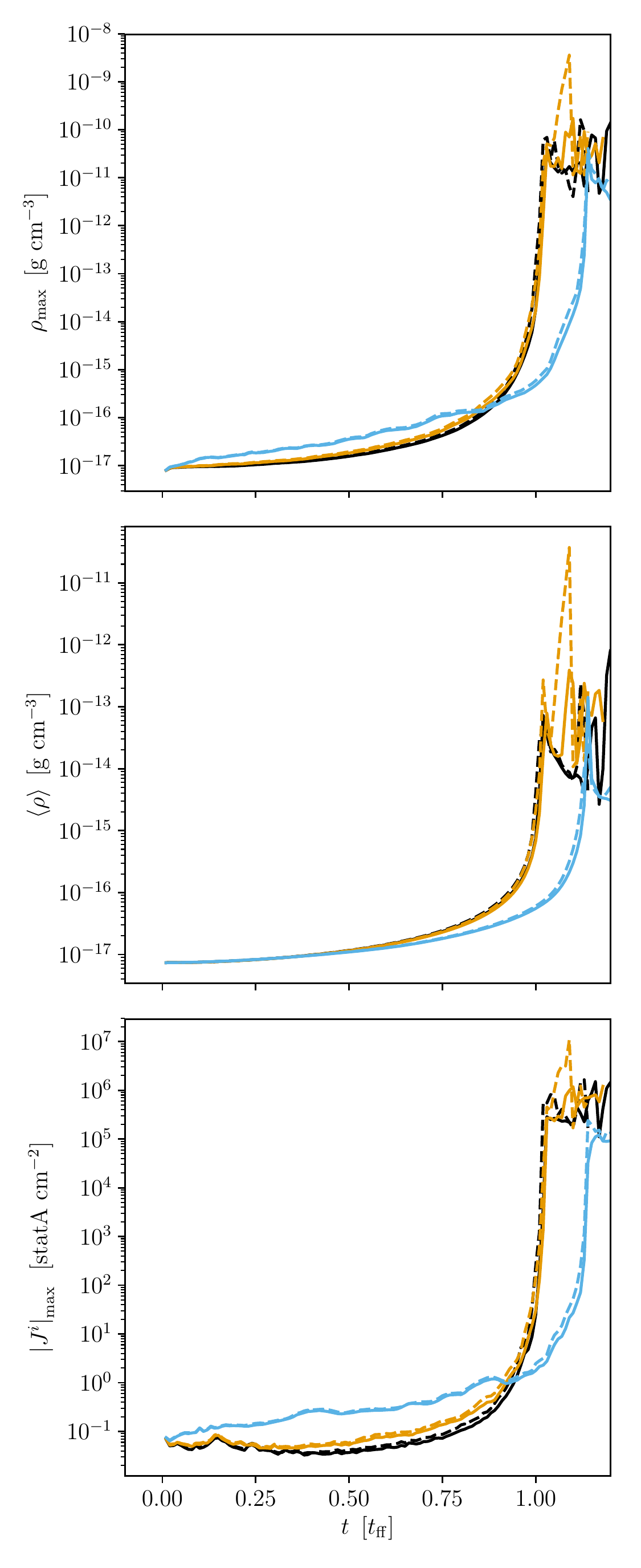}

\caption{Evolution of maximum fluid density, $\rho_\rmn{max}$, mean fluid density, $\langle\rho\rangle$, and the maximum current density, $\left|J^i\right|_\rmn{max}$, for $\Mach = 0.1$, $0.3$ and $1.0$ (shown by the black, orange and blue lines, respectively) turbulence calculations at $\mu = 5$ with both the radiative (solid lines) and barotropic (dashed lines) equations of state. The evolution until $\approx 1.00t_\rmn{ff}$ is comparable for all four sub--sonic calculations, although the addition of radiation and the increase in the Mach number to 0.3 each delay the final collapse by $\approx 0.01t_\rmn{ff}$. The two transonic calculations are also comparable until the final stage of the collapse, although there is an increased fluid density (and hence current density) in the early part of the calculation compared to the subsonic calculations due to the formation of turbulent structures.
\label{fig:MHDRTmultiplot}}
\end{figure}

\section{Dependence of the Results on the Equation of State}
\label{sec:mhdcontrarmhd}

\begin{figure*}
\centering{}
\includegraphics[width=0.84\textwidth]{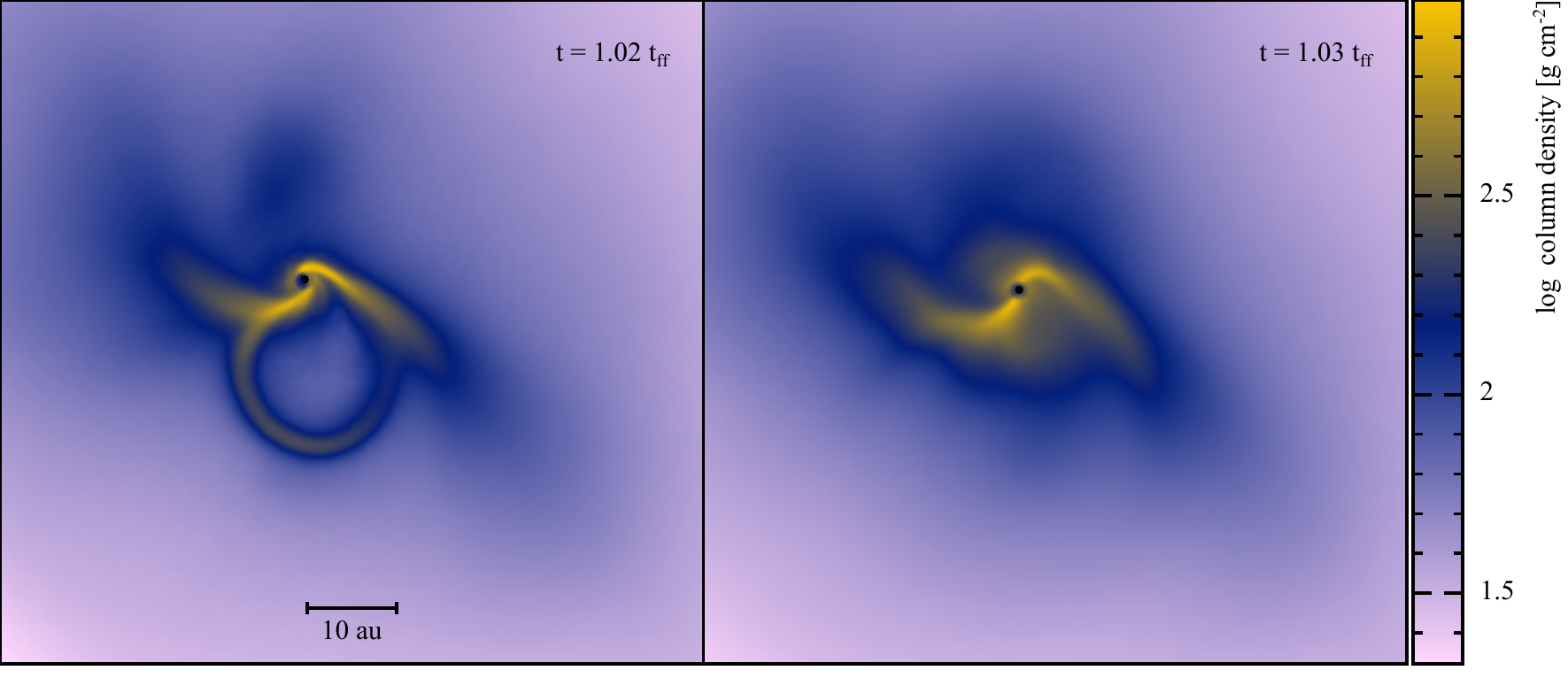}

\caption{Column density projections in the $z$--direction for the $\mu = 5$, $\Mach = 0.1$ calculations. The left panel shows the barotropic equation of state and the right panel the radiative transfer scheme. A more extensive pseudo--disc with a radius of approximately $10\text{~au}$ has been produced by the RT calculation compared to the much smaller and more unstable disc on the left panel. The barotropic equation of state produces a higher density and smaller pseudo--disc which may be more susceptible to gravitational instabilities. Alternatively, this ring--like structure may be caused by the magnetic interchange instability (see, e.g., \citealp*{1995MNRAS.275.1223S} and \citealp{1995ApJ...445..337L})
\label{fig:MHDRTM01xy}}
\end{figure*}

\begin{figure*}
\centering{}

\includegraphics[width=\textwidth]{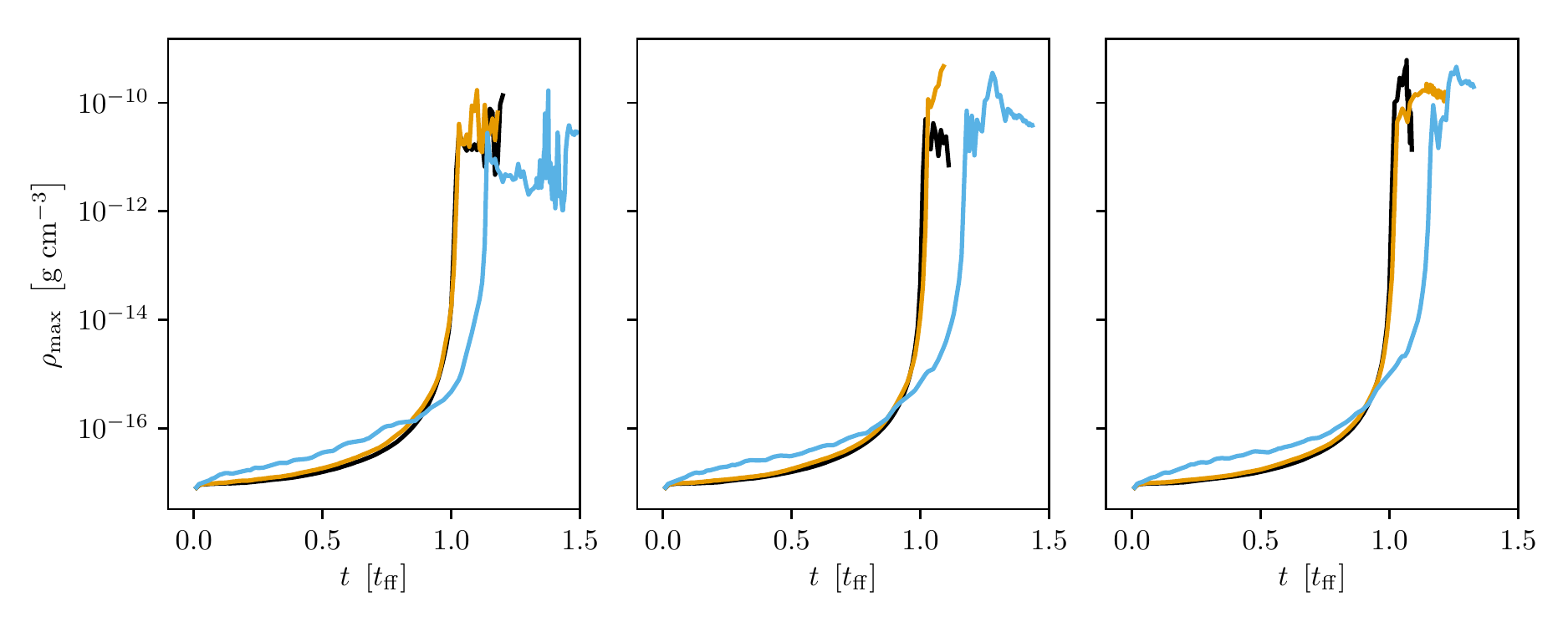}

\caption{\label{fig:maxdensevol} Evolution of the maximum fluid density for $\Mach = 0.1$ (solid black line), $\Mach = 0.3$ (short-dashed orange line), and $\Mach = 1.0$ (long-dashed purple line) initial turbulent velocity fields for an initial mass-to-flux ratio of $\mu = 5$ (left panel), $10$ (centre panel), and $20$ (left panel). For all three values of $\mu$, increasing the initial turbulence velocity delays the collapse of the molecular cloud core by providing additional energy to support the core against gravity. An effective \enquote{maximum} density is provided by the insertion of a sink particle at $\rho = 10^{-10}\udens$. 
}

\end{figure*}

\begin{figure}
\centering
\includegraphics[width=0.5\textwidth]{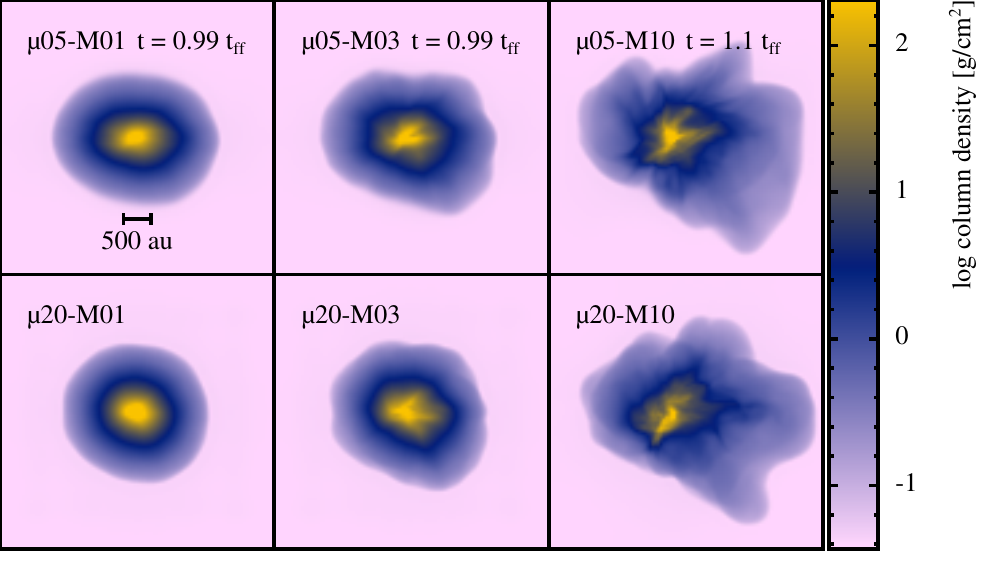}

\caption{\label{fig:oblate}Column density projections for $\Mach = 0.1$, $0.3$, $1.0$ calculations (left--to--right) with $\mu = 5$ (upper--row) and $\mu = 20$ (lower--row) showing how increasing the mass--to--flux ratio (and consequently decreasing the initial magnetic field) results in a less oblate core. This effect is less pronounced at higher turbulent Mach numbers where the structure of the core becomes dominated by the turbulent velocity field.}

\end{figure}

All the molecular cloud cores studied here are super--critical, and therefore the self-gravity of the gas will cause the core to centrally collapse. As a characteristic time, we use the free-fall time of a initially motionless (i.e. non--rotating and laminar), non-magnetised, sphere, so that for a core with mass $M_\rmn{core} = 1~\solarm$, $t_\rmn{ff} = \np{24400}~\yr$. This is in effect a lower bound on how long the core will take to collapse to \enquote{interesting} densities, and any other initial conditions, for example a large $\beta_\rmn{turb}$ will act to delay this.

Before considering the results using the radiative transfer scheme, we first consider how this differs from the barotropic equation of state. Here we show six calculations where three sets of initial conditions have been solved using both the barotropic \enquote{MHD} equation of state and the \enquote{RMHD} algorithm. The addition of RT naturally causes heating of the fluid to occur unlike the fixed equation of state, as shown by the different relationship between the fluid density and specific internal energy for the radiative equation of state and the barotropic equation of state shown in \cref{fig:trho}.
The way this translates into increased temperatures in the radiative transfer scheme is then shown by the temporal evolution of the maximum fluid temperature compared to the maximum density shown in \cref{fig:tfluid-maxdens}. We note that the density appears to reach an approximate maximum slightly between $\np{E-11}~\udens < \rho < \np{E-10}~\udens$ because of the insertion of a sink particle at (in this case) $\rho = 10^{-10}\udens$.

This increased temperature in turn promotes the formation of a larger pseudo-disc supported by the fluid pressure.
In this paper we use the same definition of pseudo--disc as in \citet{2017MNRAS.467.3324L}. This is a disc--like object that rotates at sub-Keplerian speeds around a central object (and is therefore not a \enquote{classical} accretion disc), but it is supported by the fluid pressure perpendicular to the plane of the disc against further collapse.
As expected, the calculations proceed essentially identically, with the core centrally collapsing due to self--gravity, until the fluid density has risen significantly from its initial value. \Cref{fig:MHDRTmultiplot} demonstrates how the MHD and RMHD calculations are virtually indistinguishable at this evolutionary epoch.
The $\Mach = 0.1$ and $0.3$ calculations approach a maximum density of $\approx 10^{-10}~\dens$ at $t \approx 1.02\thinspace t_\rmn{ff}$ whilst the $\Mach = 1.0$ calculation reaches a similar evolutionary point at $t \approx 1.14\thinspace t_\rmn{ff}$. After this, some divergence between the MHD and RMHD calculations can be seen. We also note that adding RT very slightly slows the collapse rate down, for instance the sink particle is inserted in the $\Mach = 0.1$ calculations $0.01\thinspace t_\rmn{ff}$ later. The additional thermal support provided by radiative feedback operates in a comparable way to how additional magnetic pressure opposes the gravitational collapse.

These results show that, even with very complicated initial conditions, the addition of radiation terms into calculations of collapsing cloud cores is important only at the shortest length scales. \Cref{fig:MHDRTM01xy} shows density projections of two pseudo--discs, one with RT and one with a barotropic equation of state. The addition of  radiation promotes stability by allowing a larger pseudo-disc, with greater pressure support to form. 
For the remainder of this paper we will consider the RMHD calculations only. 

\begin{figure*}
\centering
\includegraphics[width=\textwidth]{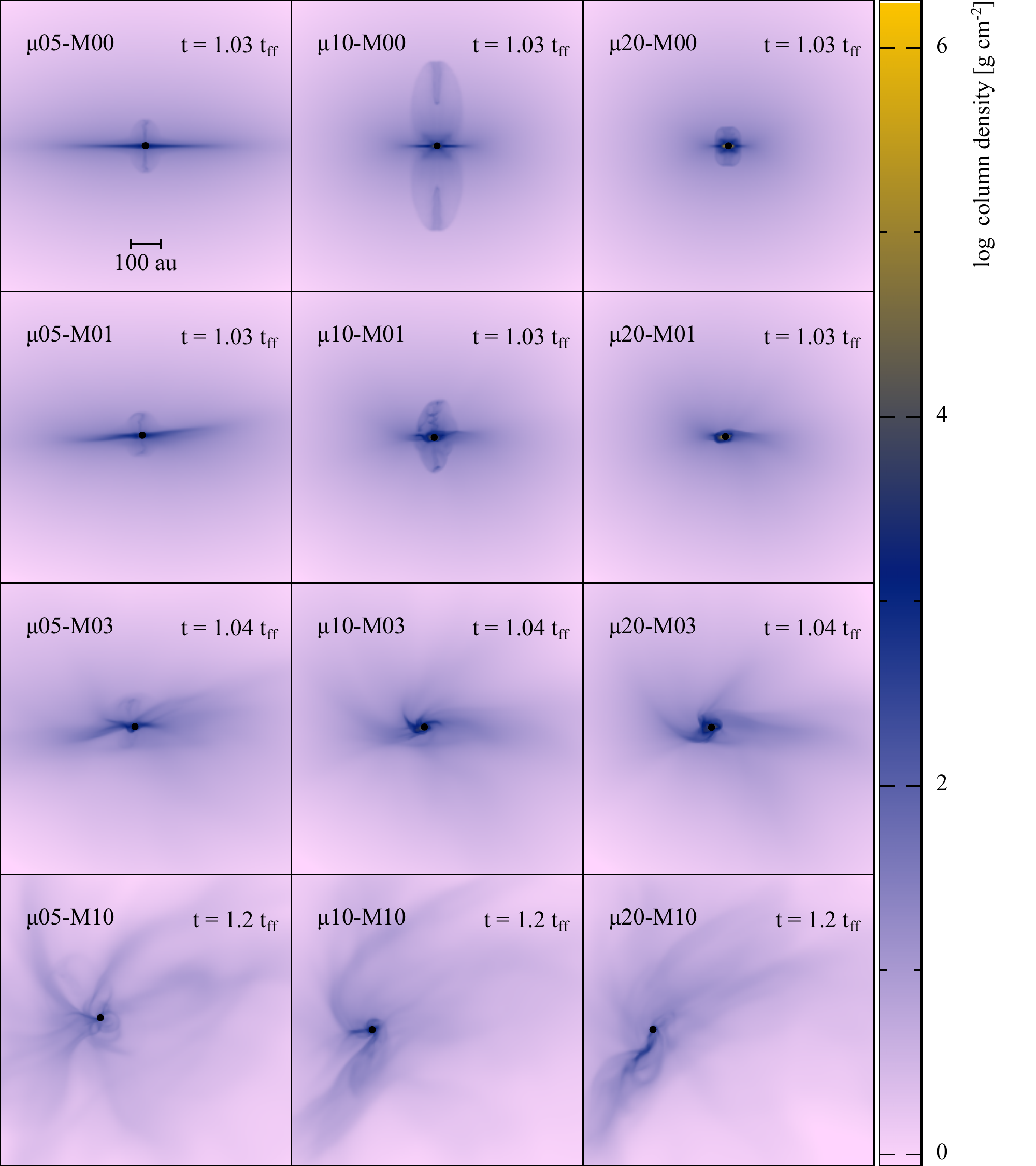}

\caption{Column density projections in the $x$-direction for twelve RMHD calculations of a collapsing molecular cloud core, the time synchronization has been chosen so that the $\mu = 5$ calculations are an equivalent time after the insertion of a sink particle (which is coterminous with a maximum density being achieved). The left--hand column has an initial magnetic field strength given by $\mu = 5$, the central column has $\mu = 10$, and the right--hand column has $\mu = 20$. The initial turbulent Mach number increases down the page, with the upper--most row being laminar calculations where $\Mach = 0$, then $\Mach = 0.1$, $0.3$ (sub--sonic cores), and $1.0$ (transonic) in the bottom row. Bipolar outflows are only obvious for $\mu \leq 10$ coupled with $\Mach \leq 0.1$ although the $\mu = 5$, $\Mach = 0.3$ calculation has a weak outflow.
\label{fig:allmus}}
\end{figure*}

\section{The Initial Phase of the Collapse}
\label{sec:first}

\Cref{fig:maxdensevol} shows the evolution of the maximum density for three turbulent Mach numbers ($\Mach = 0.1$, $0.3$, and $1.0$) across three mass-to-flux ratios ($\mu = 5$, $10$, $20$), which correspond to models $\upmu$05--M01(R), $\upmu$05--M03(R), $\upmu$05--M10(R), $\upmu$10--M01, $\upmu$10--M03, $\upmu$10--M10, $\upmu$20--M01, $\upmu$20--M03, and $\upmu$20--M10 in \cref{tbl:initialcond}. 
As discussed in relation to \cref{fig:tfluid-maxdens}, the fluid appears to reach a maximum density because of the insertion of a sink particle.
Here we see that the degree of turbulent support has a vastly greater influence on the collapse time of the core. Magnetic pressure support is extremely weak until the core reaches the highest densities and consequently decreasing the mass-to-flux ratio from, for example, $10$ to $5$ has a much smaller impact. We observed in \mylb and earlier in \mylbp that the initial field strength and geometry has an effect on the collapsing core, and we again see this effect here. There is (effectively) no magnetic pressure support along the field lines, and this is consequently the preferential direction for fluid to graviationally collapse. \Cref{fig:oblate} shows the evolution of three $\mu = 5$ cores and three $\mu = 20$ cores with $\rho_\rmn{max}\approx 10^{-11}\udens$ where the characteristic \enquote{oblate spheroid} shape proportional to the initial field strength can be clearly seen when the Mach number $\leq 0.3$. We obtain a comparable range of ellipticities (i.e. axis ratios) to the laminar cores presented in \mylb with values of 0.6, 0.64, for $\mu = 5$, $\Mach = 0.1$ and $0.3$ and 0.29, and 0.33 for $\mu = 20$ respectively, compared to 0.66 and 0.24 for a laminar core. Although the transonic $\Mach = 1$ cores are clearly somewhat oblate, it is not possible to demonstrate (due to the complex structure) that the $\mu = 20$ core has a lower oblateness than for $\mu = 5$. In any event, in this transonic regime the turbulent kinetic energy --- and hence the filament-like structures produced ---  has a much greater effect on the initial collapse than the magnetic field (\cf \cref{fig:maxdensevol}). 

Once the core reaches densities of $\approx 10^{-10}~\udens$, the magnetic field becomes a significant contributor to the subsequent evolution of the system.
We do not include physical dissipative magnetic effects such as Ohmic resistivity or ambipolar diffusion. At densities in excess of $\approx 10^{-12}~\udens$ magnetic dissipation can become important \citep{1976PASJ...28..355N,2002ApJ...573..199N}, however, in all the calculations presented here any substantial region with this high density will either be rapidly accreted or will form a sink particle. Consequently, neglecting these effects should not significantly impact the overall evolution of the pseudo--discs formed or any outflows produced.
In addition, because of the radiative transfer scheme, radiation effects become important, as discussed earlier (see, e.g., \cref{fig:tfluid-maxdens}). 

In summary, at high Mach numbers the turbulent velocity field dominates the initial phase of the evolution. RT becomes important when the density exceeds $\approx 10^{-13}~\udens$. MHD effects in the initial phase are limited to changing the oblateness of the core, more complicated magnetic effects only occuring when the density is higher than about $10^{-11}~\udens$. At lower Mach numbers, the turbulent velocity field instead acts to perturb and disrupt the uniformity of the collapse. This highlights a small difficulty in using the ideal free--fall time as a dimensionless temporal unit: the number of free--fall times at which the core will reach an \enquote{interesting} density will vary depending on the sources of support against gravitational collapse present, for example a $\Mach = 1$ core will reach these densities $\sim\nicefrac{1}{5}~\tff$ later than a laminar core. However, this unit is still useful for providing a dimensionless comparison between differing mass and volume cores. Consequently, in the following sections we will in general synchronize the calculations so that a comparable evolutionary epoch is discussed, rather than a fixed $\tff$. 

\section{Shaken: varying the initial Mach number}
\label{sec:shaken}

\begin{figure}
\centering{}
\includegraphics[width=0.4\textwidth]{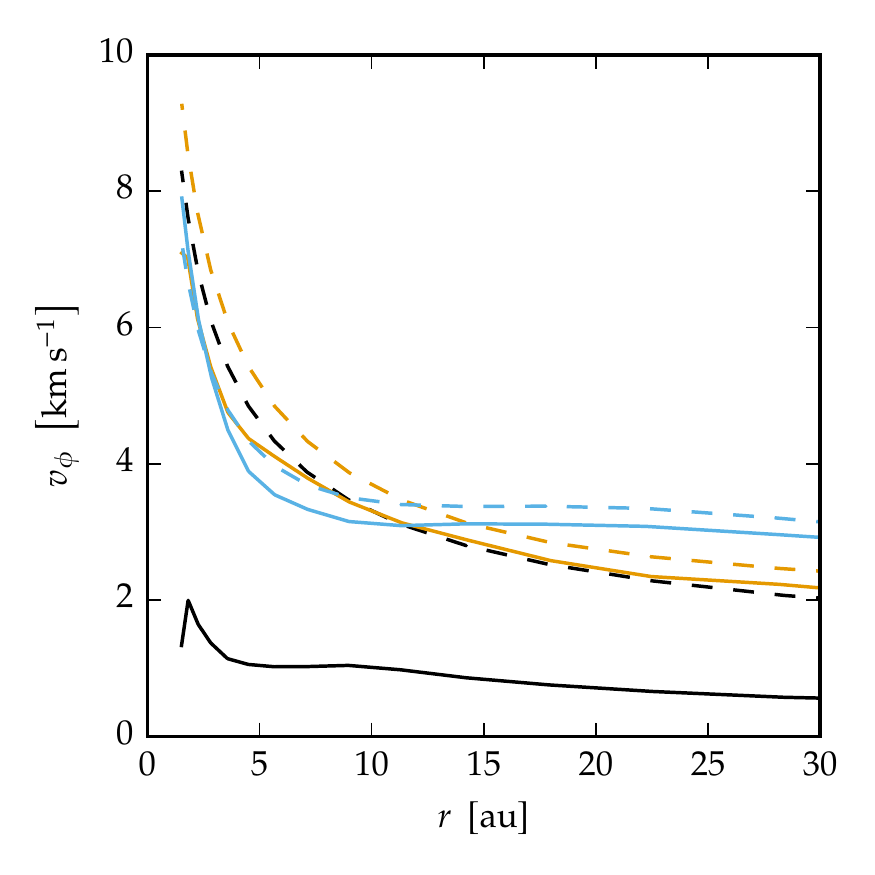}\\
\includegraphics[width=0.4\textwidth]{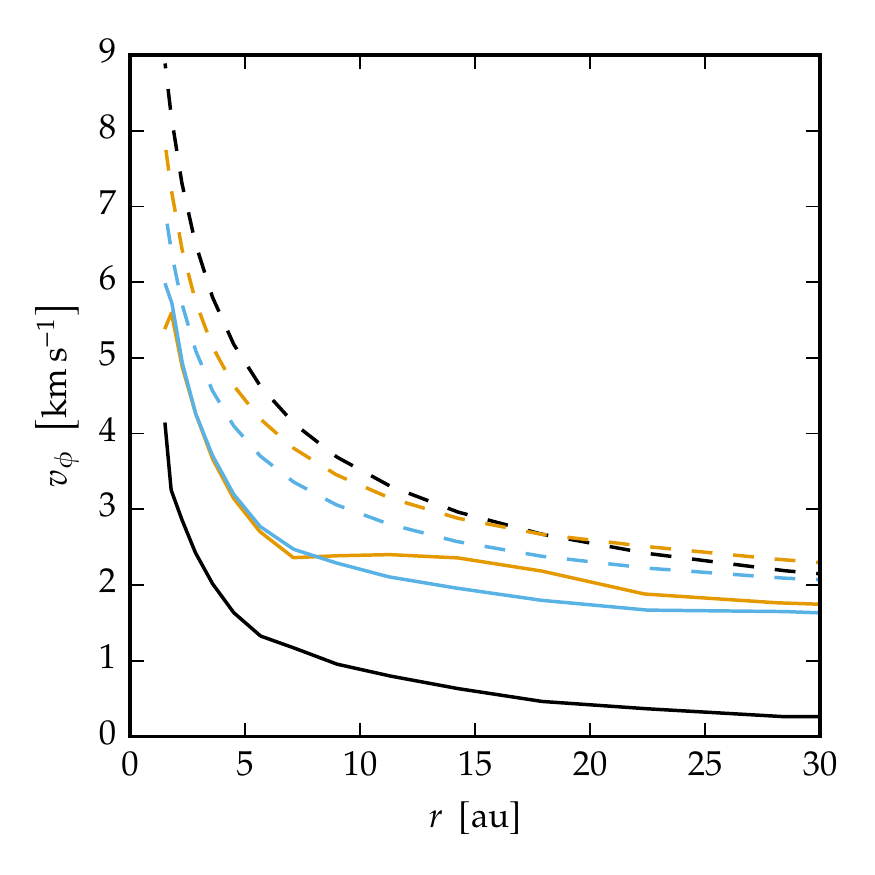}

\caption{Radially averaged azimuthal velocity as a function of radius from the sink--particle for the three $\mu = 5$, $10$, and $20$ calculations (black, orange and blue solid lines respectively) compared to the Keplerian velocity at that radius (dashed lines). The upper panel is for $\Mach = 0.1$ and the lower panel is for $\Mach = 0.3$. 
The pseudo--discs become progressively further from Keplerian as $\mu$ decreasing (i.e. as the field strength increases); in addition, for weaker fields increasing $\Mach$ also results in a more sub--Keplerian rotation profile.
 \label{fig:radvel}}
\end{figure}

\Cref{fig:allmus} shows the effect of increasing the initial turbulent velocity field for all three mass-to-flux ratios. A clear pattern emerges: firstly as in our previous laminar work, $\mu = 20$ calculations are unable to produce a bipolar outflow --- which we see in the third column of \cref{fig:allmus}. Secondly, that increasing the Mach number disrupts the system and inhibits the formation of a jet, as seen in the third and fourth rows of that figure. Throughout this paper we adopt a similar classification to our earlier works. The term `outflow' is used to describe \textit{any} fluid propelled away from the protostellar core, whereas the term `jet' is reserved for defined structures exhibiting at least some degree of collimation. A rotating pseudo--disc is a necessary precursor to the formation of a bipolar jet (see, \eg, \citealp*{2003MNRAS.339.1223P}, \citealp{2010MNRAS.409L..39C}, \citealp*{2012A&A...543A.128J}, \citealp{2015MNRAS.451.4807L}) and an initially turbulent core acts to prevent the formation of this disc. \Cref{fig:allmus} demonstrates this effect: at $\Mach = 0.1$ the $\mu = 5$ and $10$ calculations clearly exhibit disc structures similar to, though less uniform, those produced by laminar calculations, which are increasingly disrupted as the initial turbulent Mach number is increased. This effect is also present (albeit without the bipolar jets) for $\mu = 20$ \citep[see also][]{2012A&A...543A.128J}. \Cref{fig:radvel} compares the radially averaged azimuthal velocity of the three $\Mach = 0.1$ discs. We do not include more turbulent cores on this figure because the highly disrupted nature of the discs results in an impossibly confused graph.  
These results are broadly comparable with the conclusions of \citet{2017arXiv170309139M}. 

We also observe that adding radiative transfer to our laminar calculations does not markedly affect the shape or velocity of the outflow produced compared to the barotropic results obtained in \citet{2017MNRAS.467.3324L}. This concurs with our observations in \cref{sec:mhdcontrarmhd} \citep[and also with][]{2014MNRAS.437...77B} that adding a radiative transfer treatment is a small overall effect, operating principally to produce a larger and warmer disc. 

\begin{figure*}
\centering{}
\includegraphics[width=0.3333333333\textwidth]{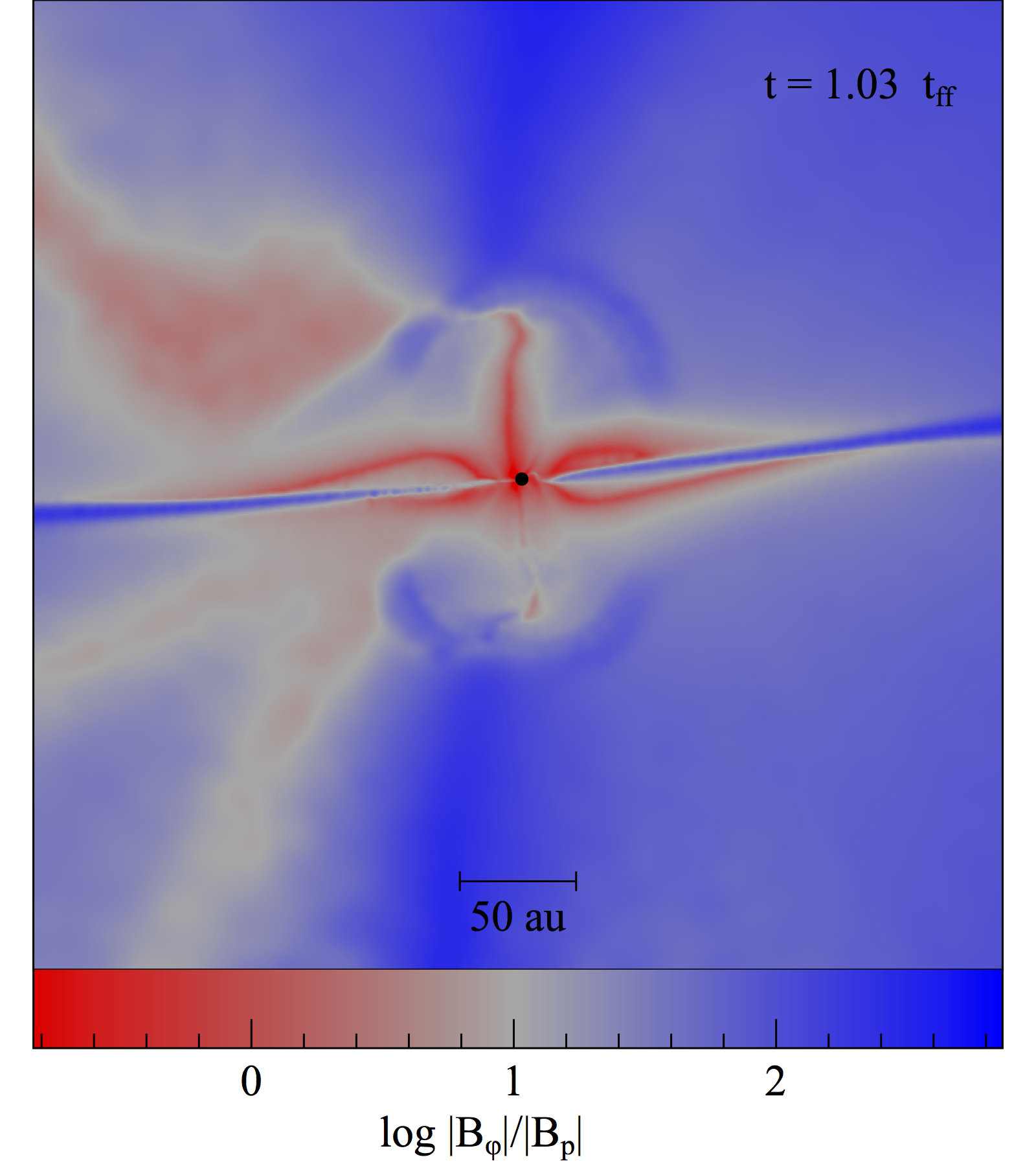}\includegraphics[width=0.3333333333\textwidth]{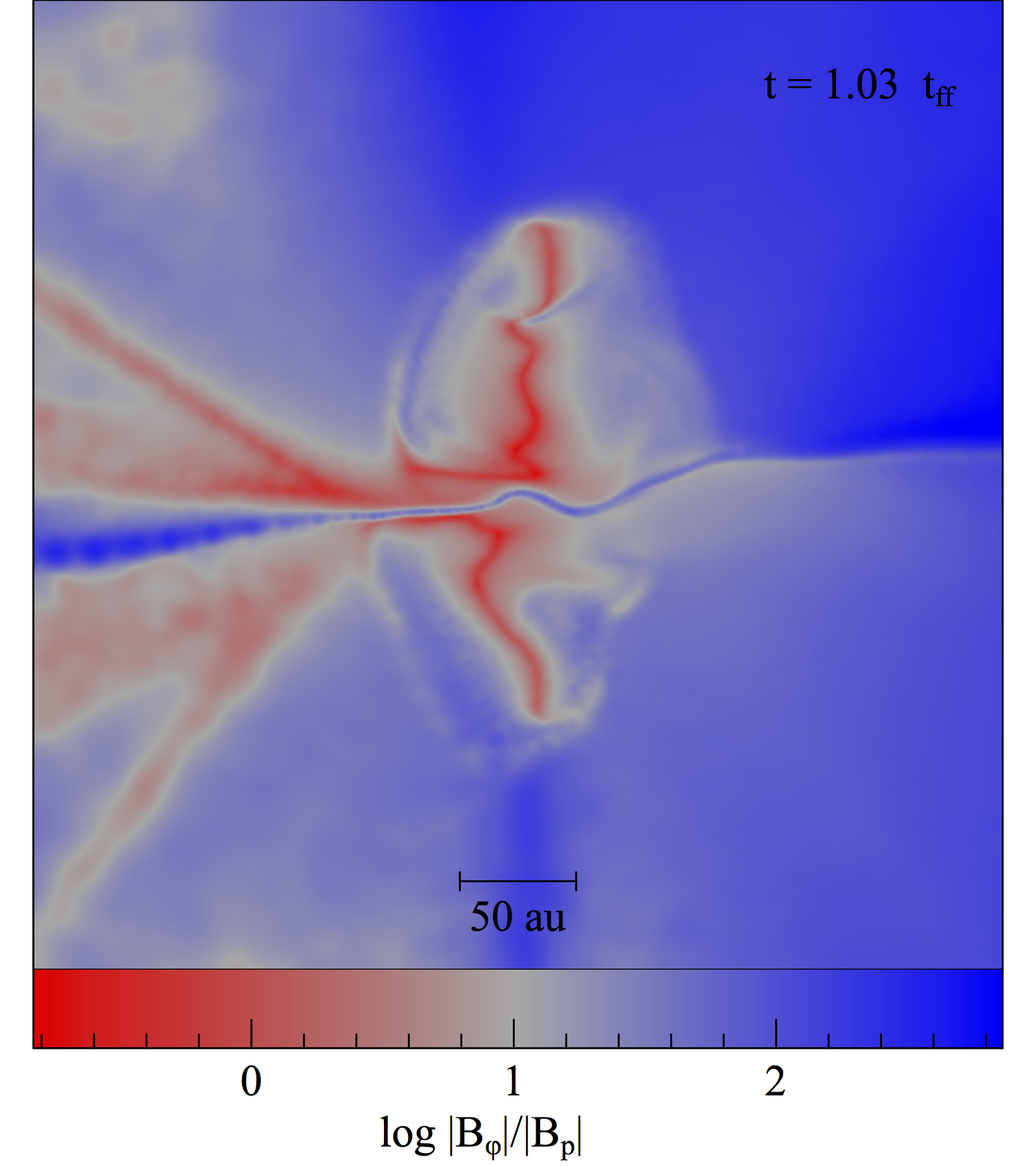}\includegraphics[width=0.3333333333\textwidth]{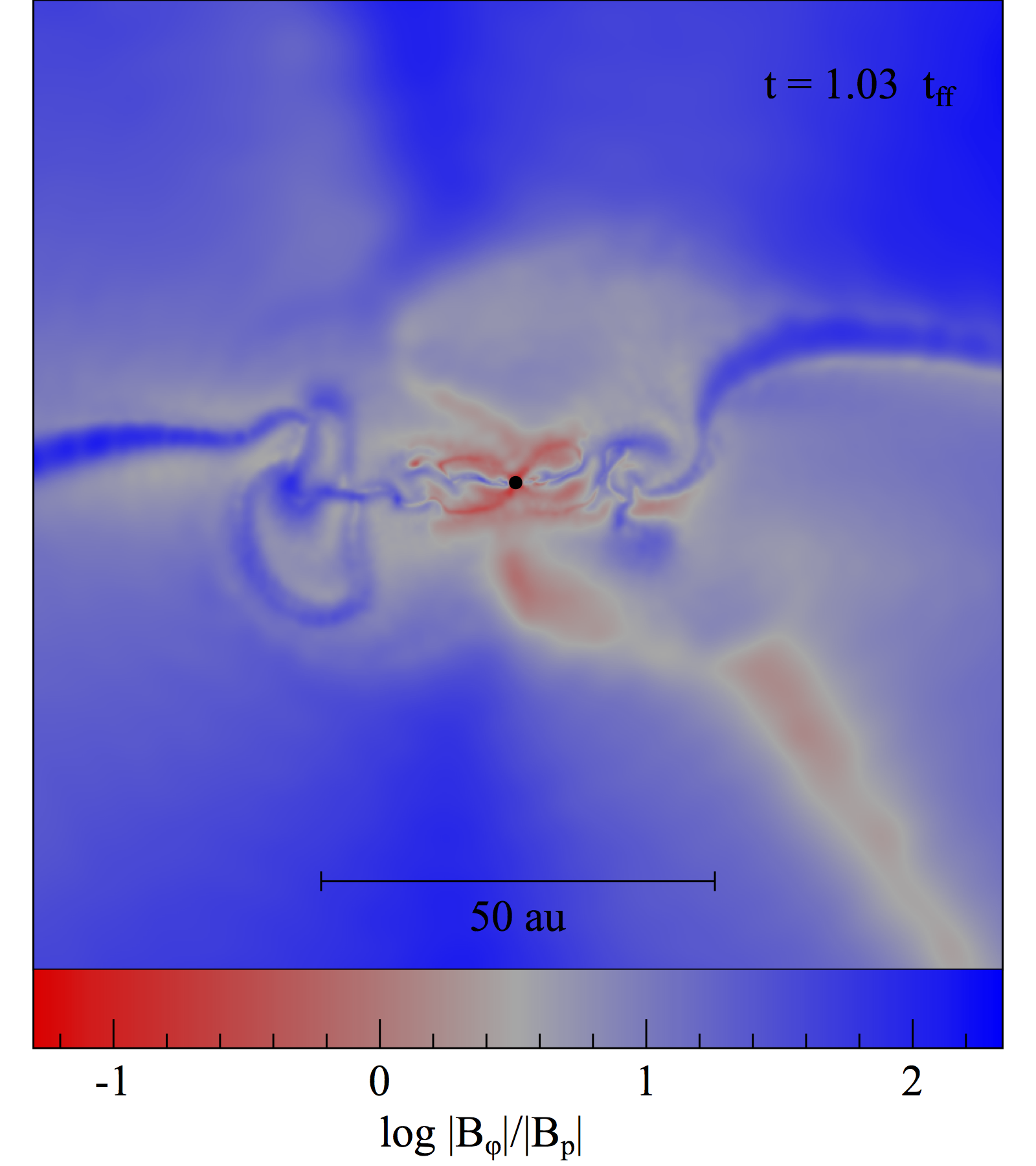}
\includegraphics[width=0.3333333333\textwidth]{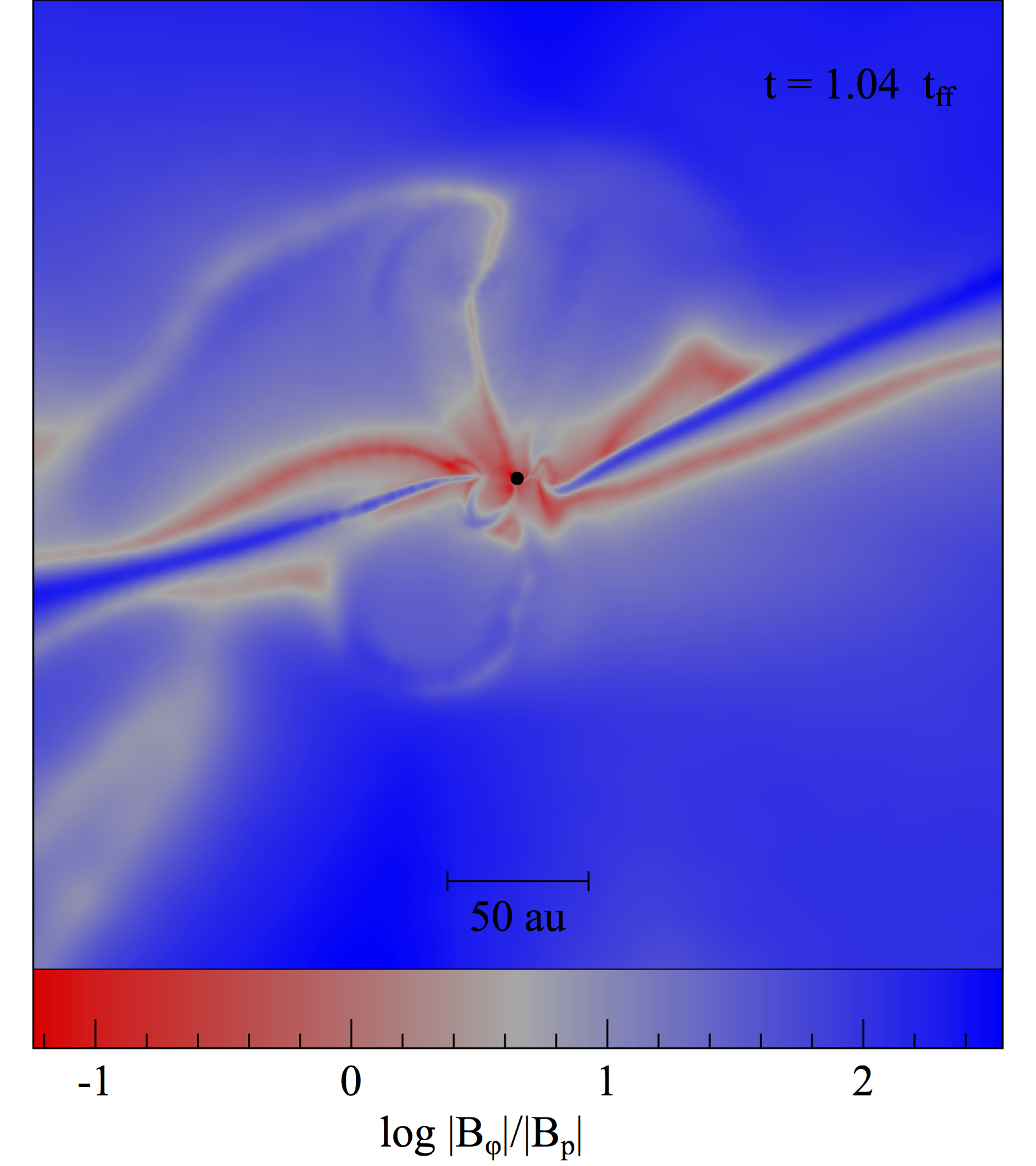}\includegraphics[width=0.3333333333\textwidth]{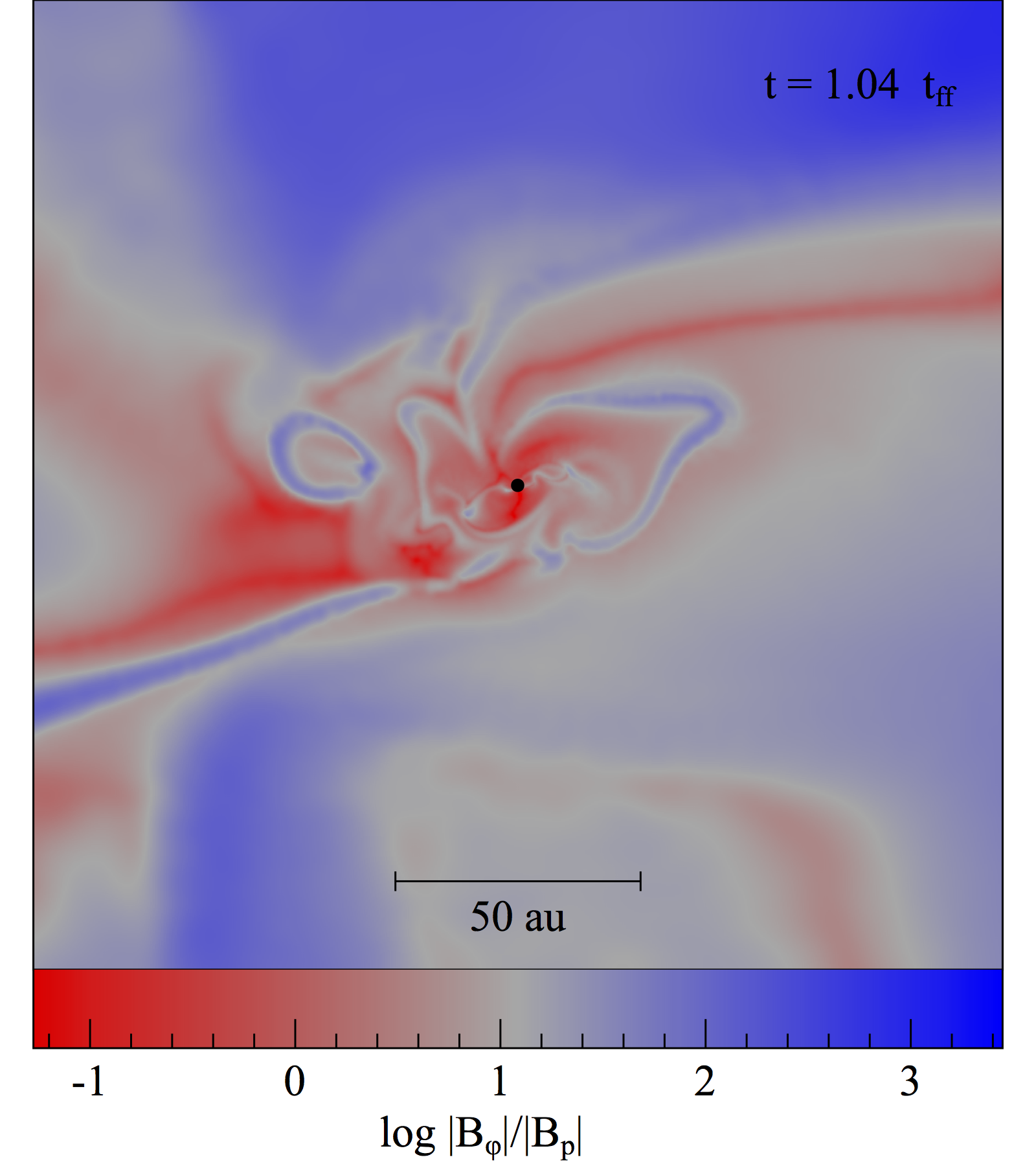}\includegraphics[width=0.3333333333\textwidth]{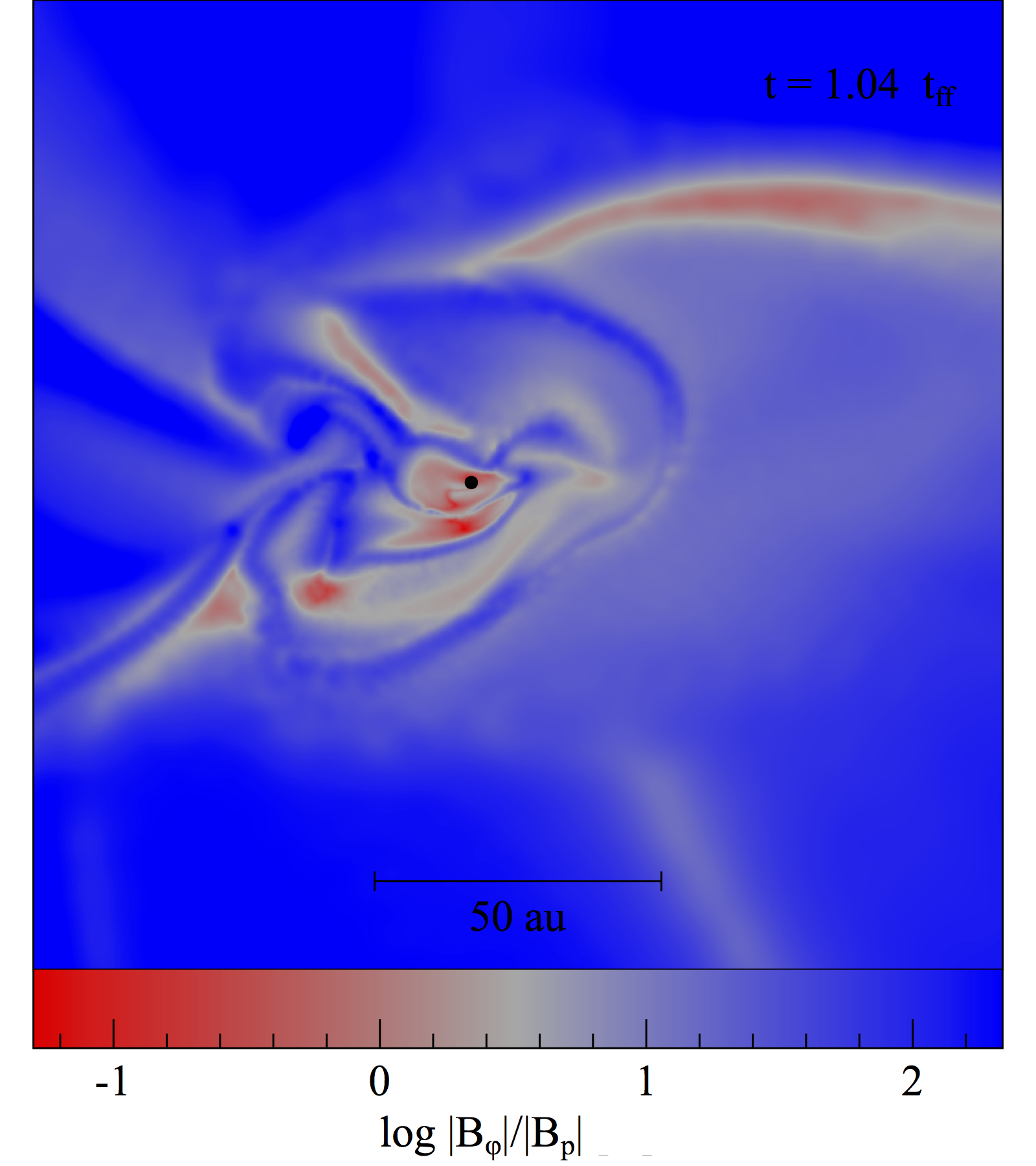}
\includegraphics[width=0.3333333333\textwidth]{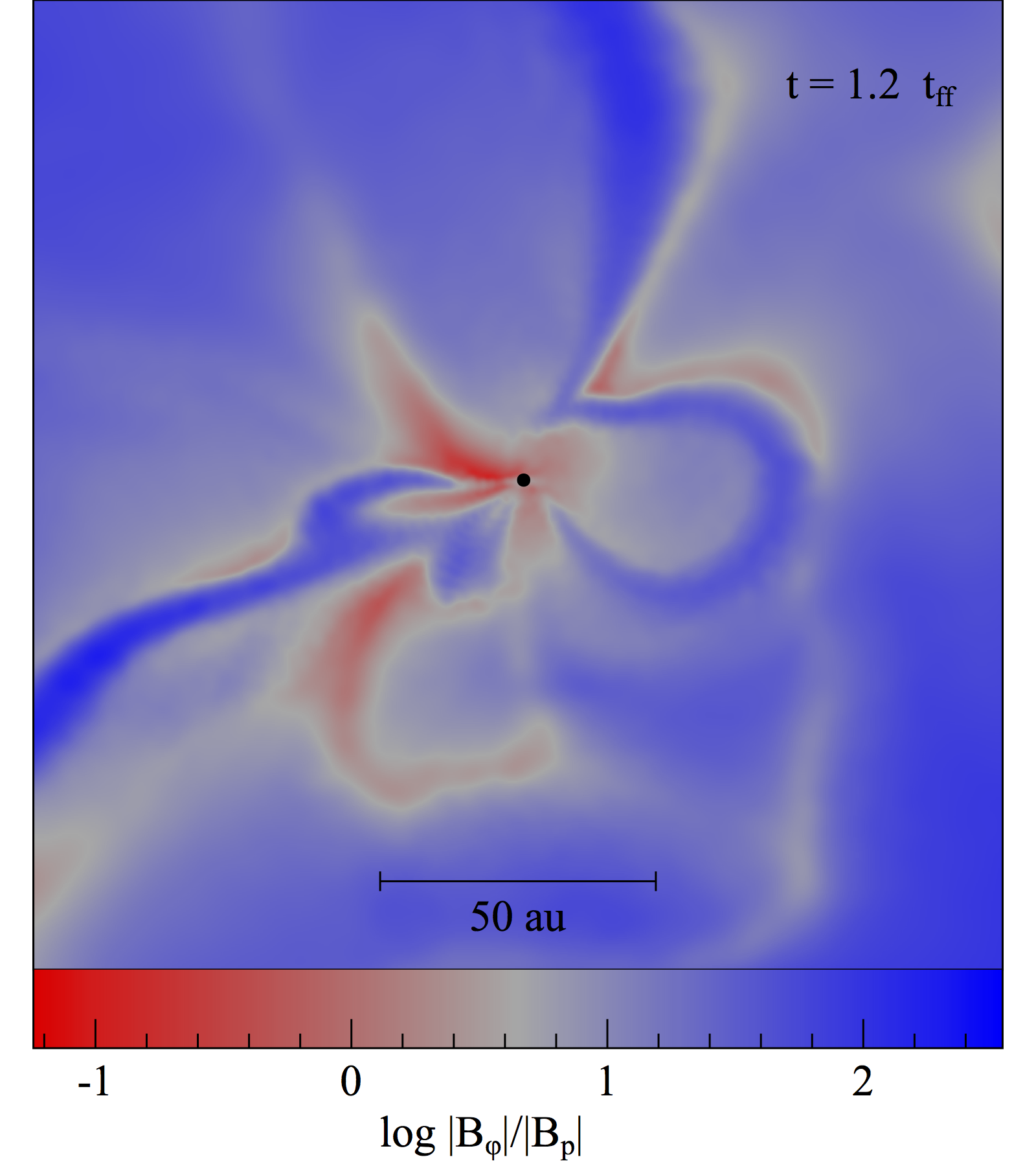}\includegraphics[width=0.3333333333\textwidth]{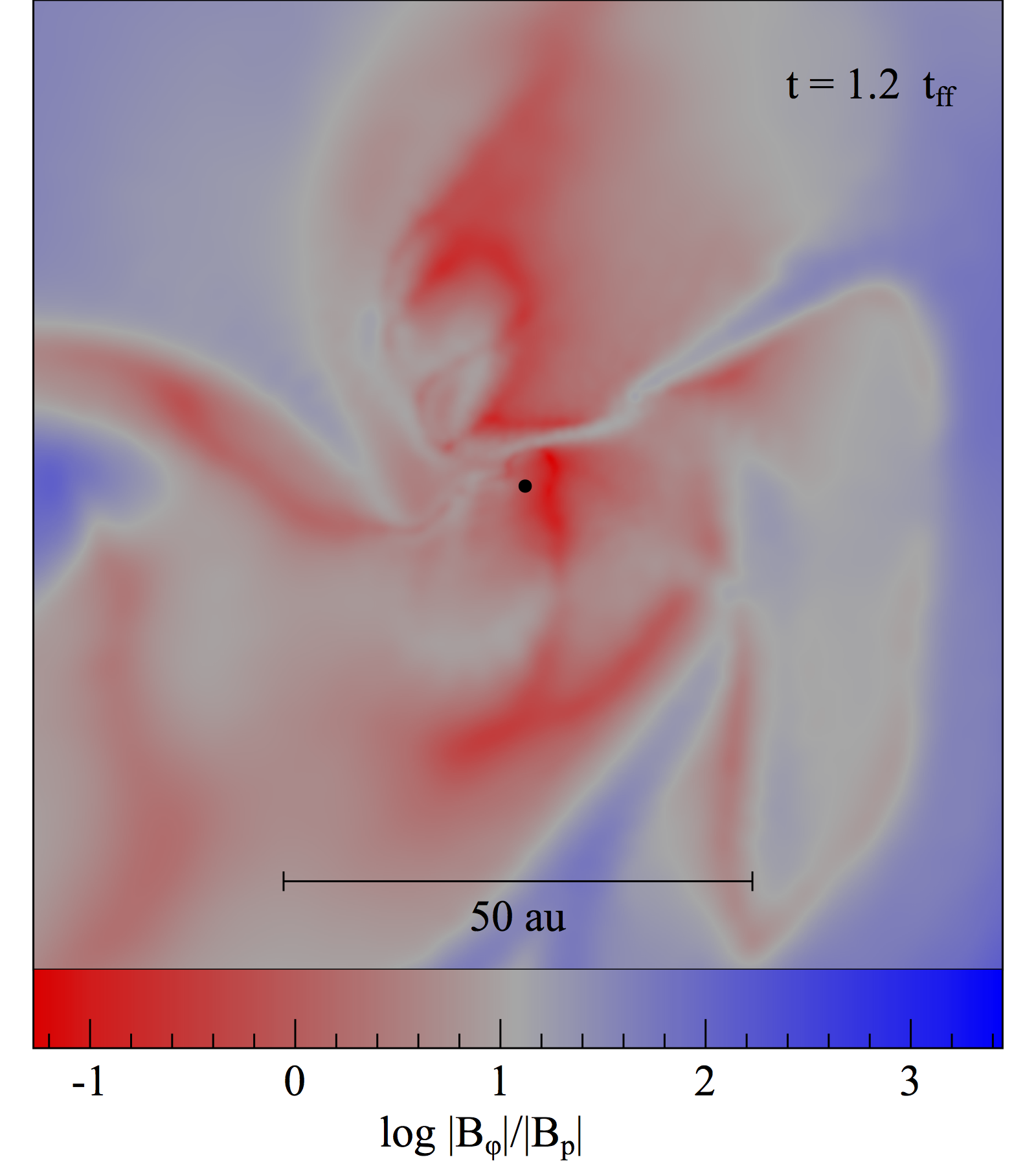}\includegraphics[width=0.3333333333\textwidth]{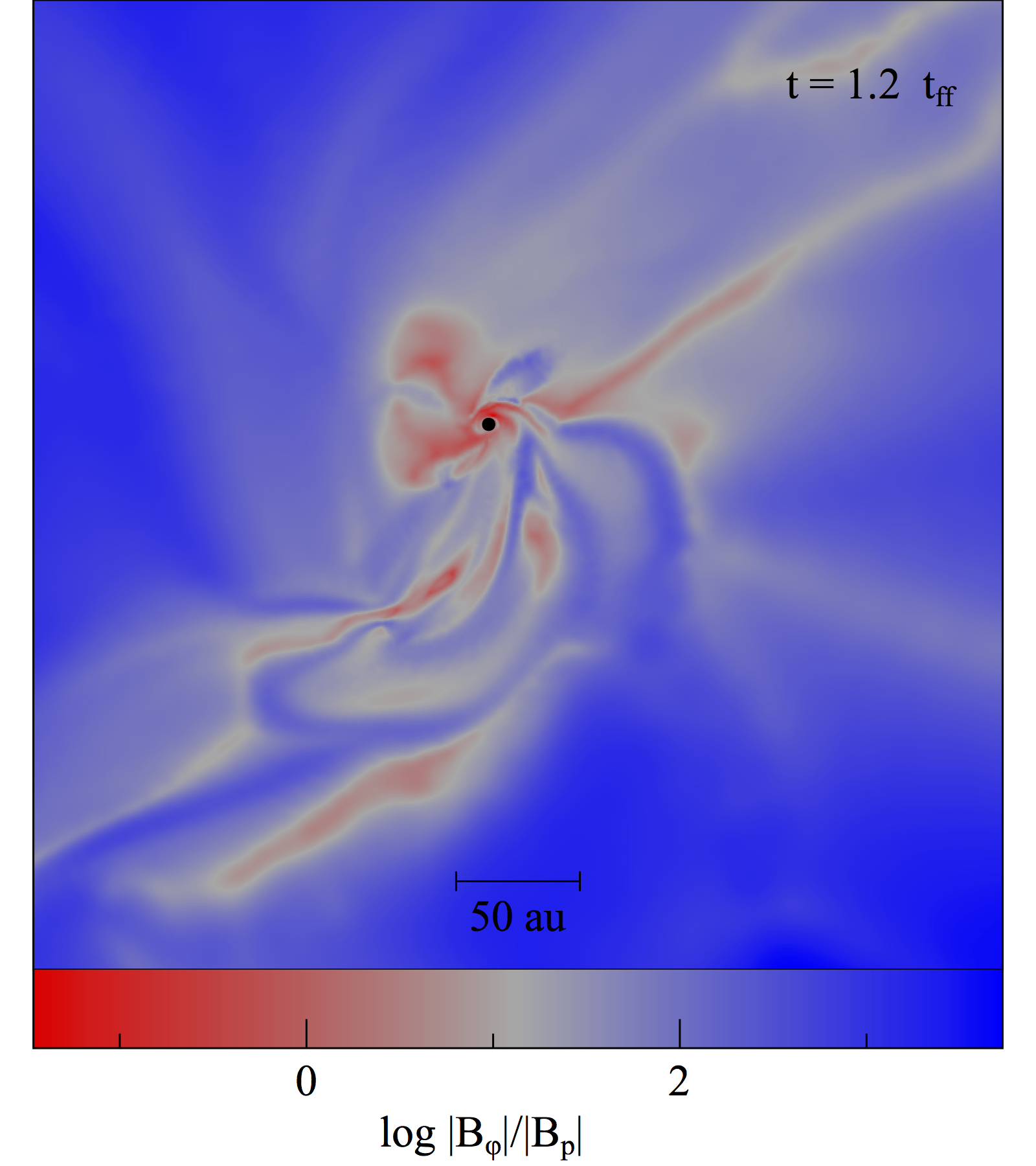}

\caption{Transverse sections, in a $z$-$x$ plane through the sink particle of the ratio of the poloidal magnetic field component to the toroidal component for all nine turbulent calculations presented in \cref{fig:allmus}. The sections are presented in the same order as in that figure, so that the top middle and bottom rows are $\mu = 5$, $10$, and $20$ respectively and the left, centre and right columns are $\Mach = 0.1$, $0.3$, and $1.0$ respectively. Red indicates regions where the poloidal field component (\cf \cref{eqn:polB}) is dominant and conversely blue indicates regions where the toroidal field (\cf \cref{eqn:torB}) is dominant. Material is removed from the region around the sink--particle by the poloidal field. This process, however, is more efficient and results in a collimated jet --- in particular as seen in the top--left panel --- when the toroidal field is able to wind up the material as it is being lifted away from a pseudo-disc. 
\label{fig:shakenTP}}
\end{figure*}

\begin{figure*}
\centering{}
\includegraphics{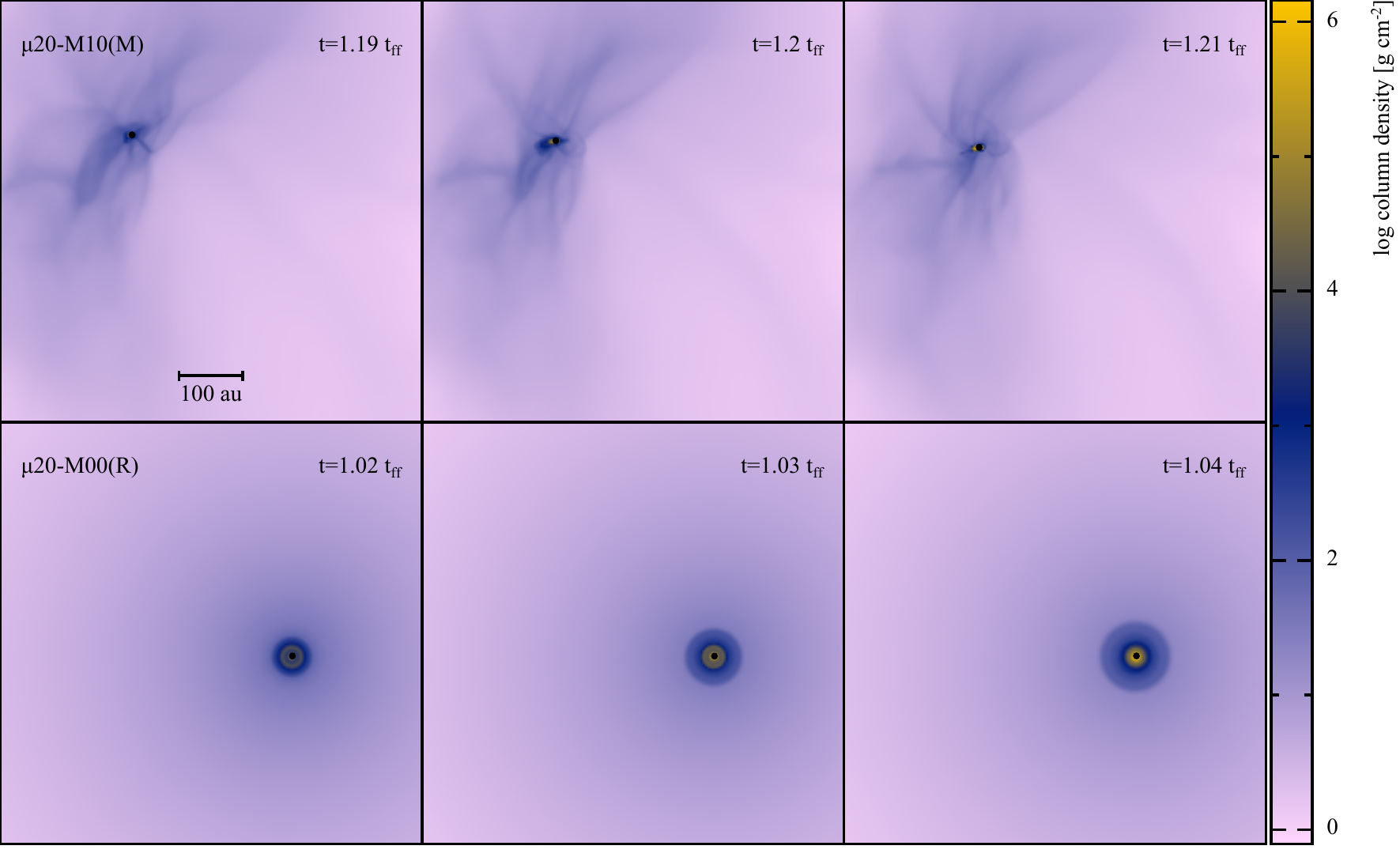}

\caption{\label{fig:mu20_Fragments}Column density projections in the $y$--direction for $\upmu$20--M10(M) (upper row) and $\upmu$20-M00(R) (lower row) at a comparable evolutionary epoch after a sink particle is inserted. Neither calculation has fragmented, unlike the $\mu = 20$ calculations in \citet{2017MNRAS.467.3324L}, demonstrating that adding \textit{either} turbulence (see the upper row) or RT (lower row) is sufficent to prevent this fragmentation mode.}
\end{figure*}

\begin{figure*}
\centering{}
\includegraphics[width=\textwidth]{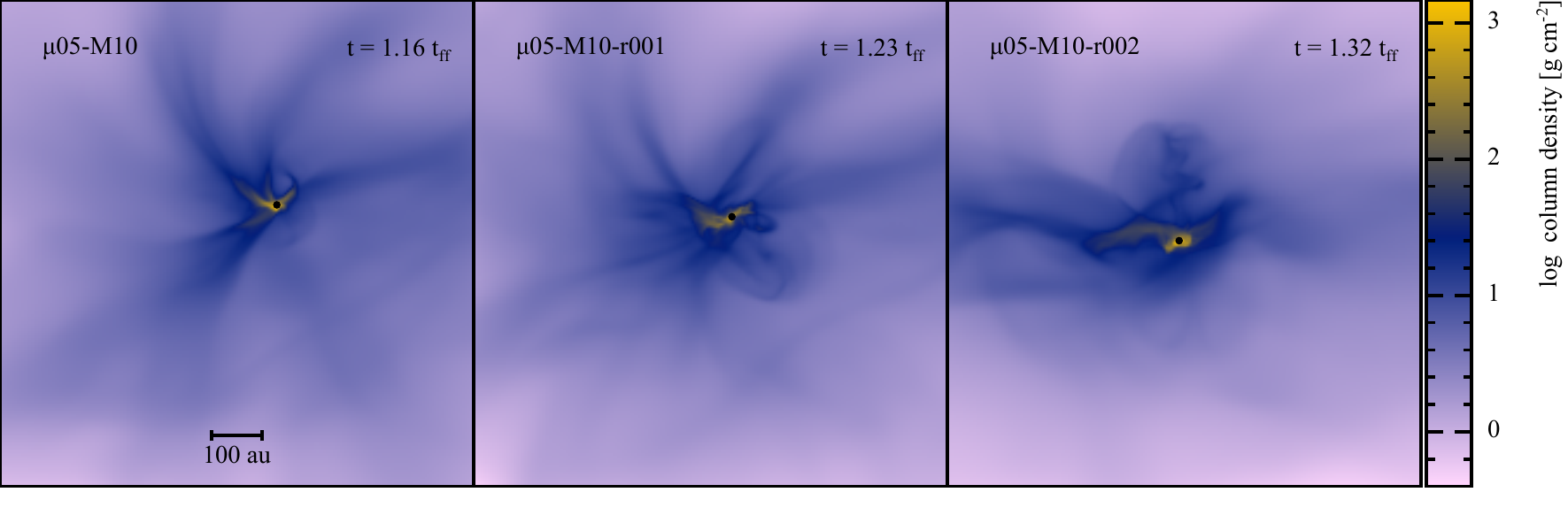}
\includegraphics[width=\textwidth]{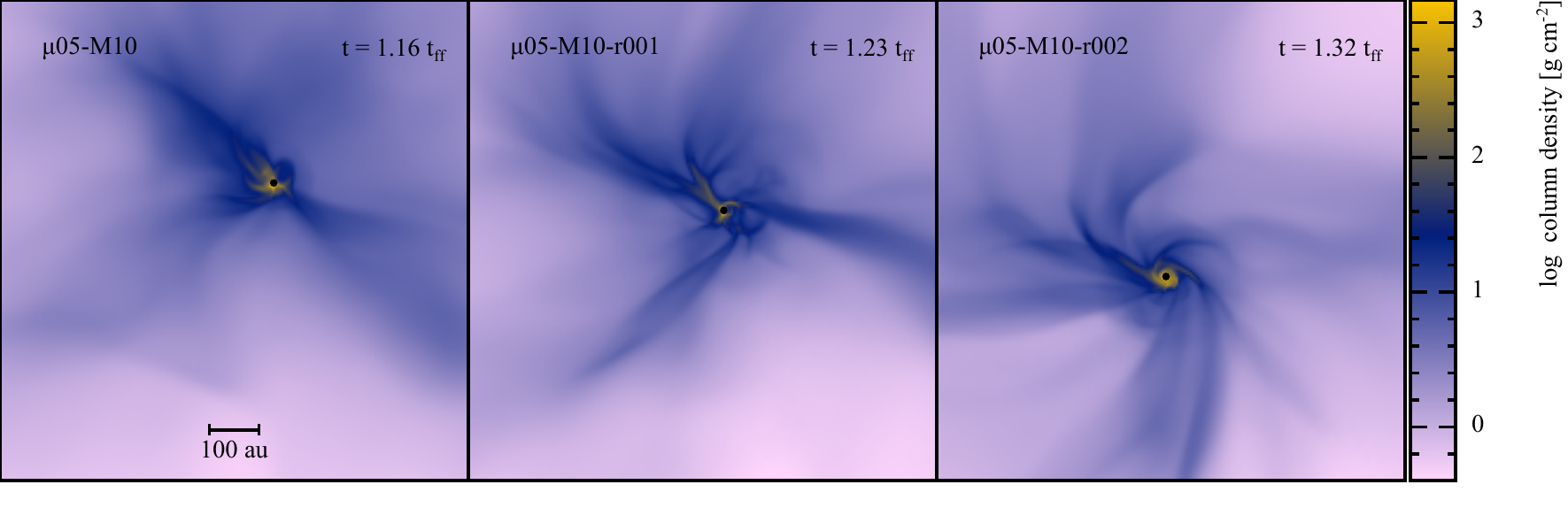}

\caption{Column density projections in the $x$-direction (upper row) and $z-$direction  (lower row) for three RMHD calculations of a collapsing molecular cloud core, all with $\mu = 5$ and an initial Mach number of 1.0. The left panel is the same calculation presented in \cref{sec:shaken}, shown here again as a comparison, which has $\beta_\rmn{rot} = 0.005$. The centre and right panels then have $\beta_\rmn{rot} = 0.01$ and $0.02$, \ie double and quadruple, respectively. 
\label{fig:stirred}}
\end{figure*}

\begin{figure*}
\centering{}
\includegraphics[width=\textwidth]{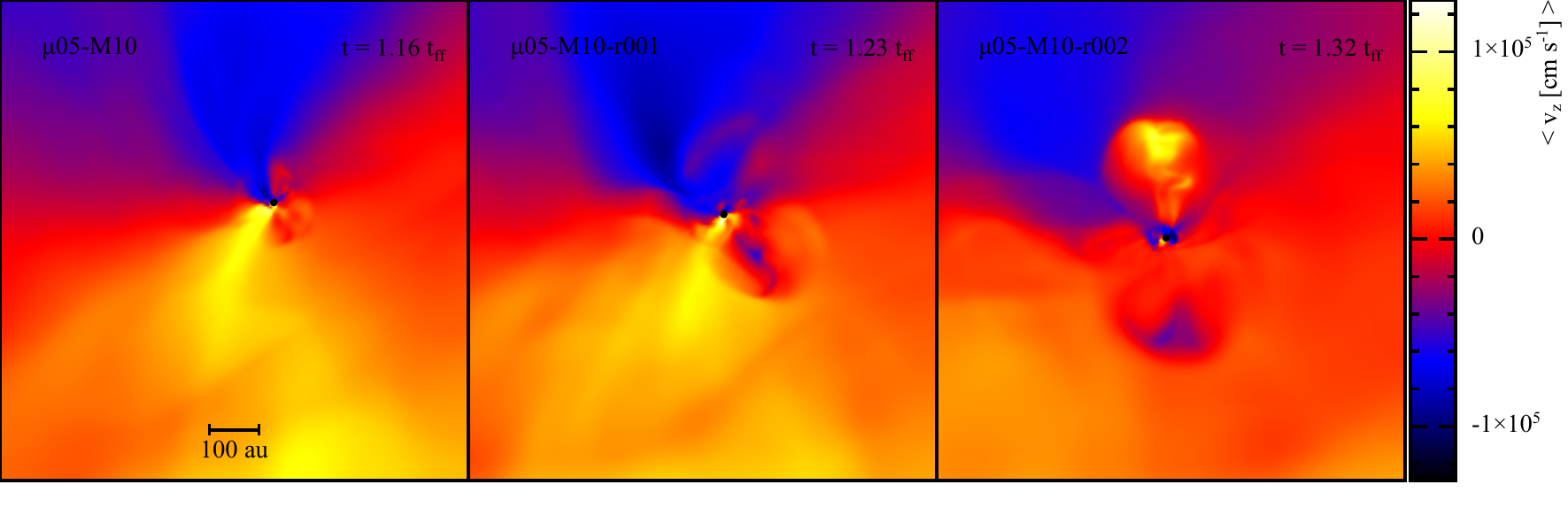}

\caption{Density-weighted average of the $z-$component of the velocity field for the three calculations presented in \cref{fig:stirred}, \ie where the left panel, centre panel and right panel have $\beta_\rmn{rot} = 0.005$, $0.01$, and $0.02$ respectively. The signature of an outflow, a region with material moving the \enquote{wrong way} is clearly visible in the right--hand panel, as opposed to the highly disrupted calculations presented in the other two panels.
\label{fig:stirredvz}}
\end{figure*}

\begin{figure*}
\centering{}

\includegraphics[width=0.3333333333\textwidth]{plots/TP/1_3.png}\includegraphics[width=0.3333333333\textwidth]{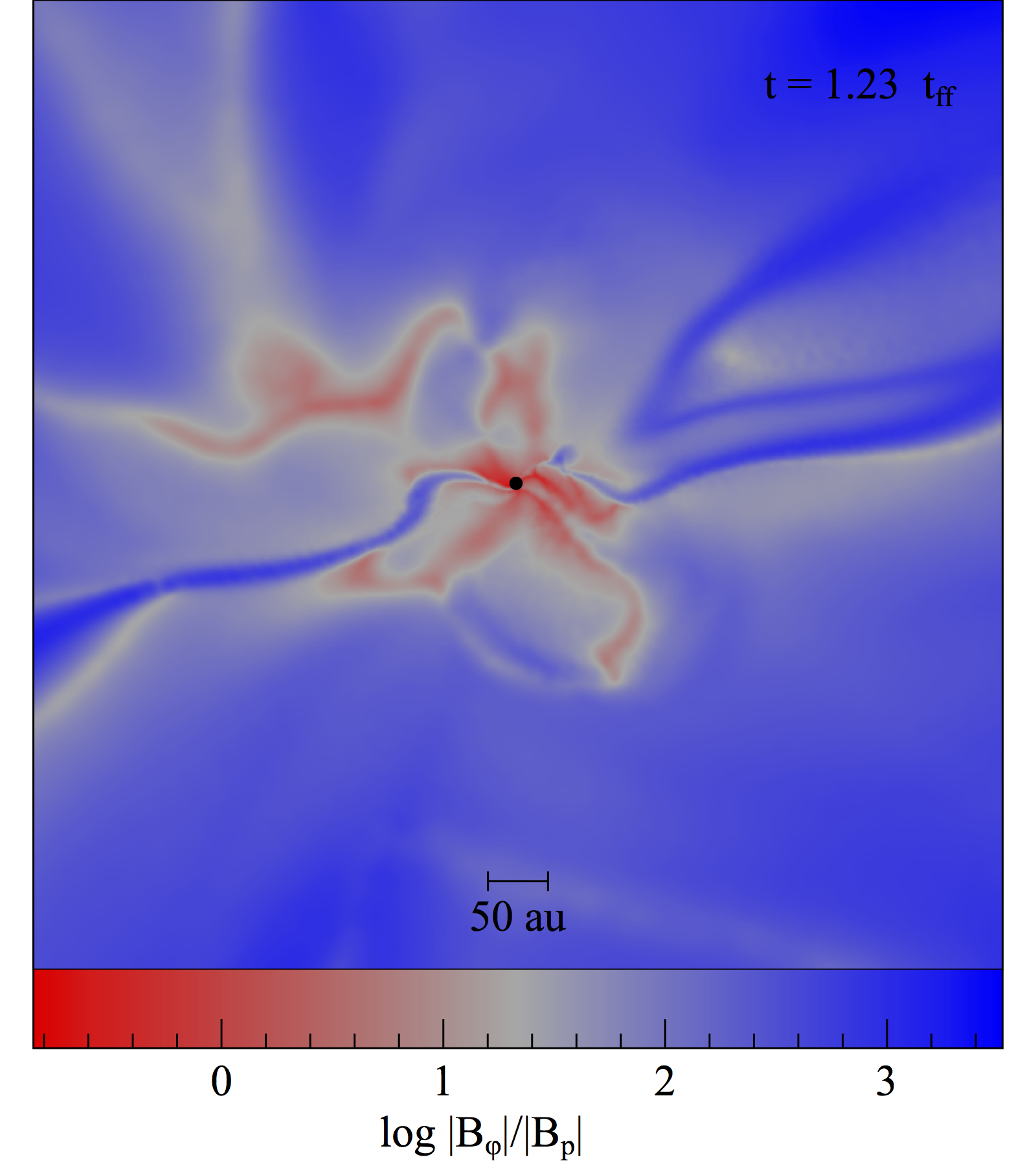}\includegraphics[width=0.3333333333\textwidth]{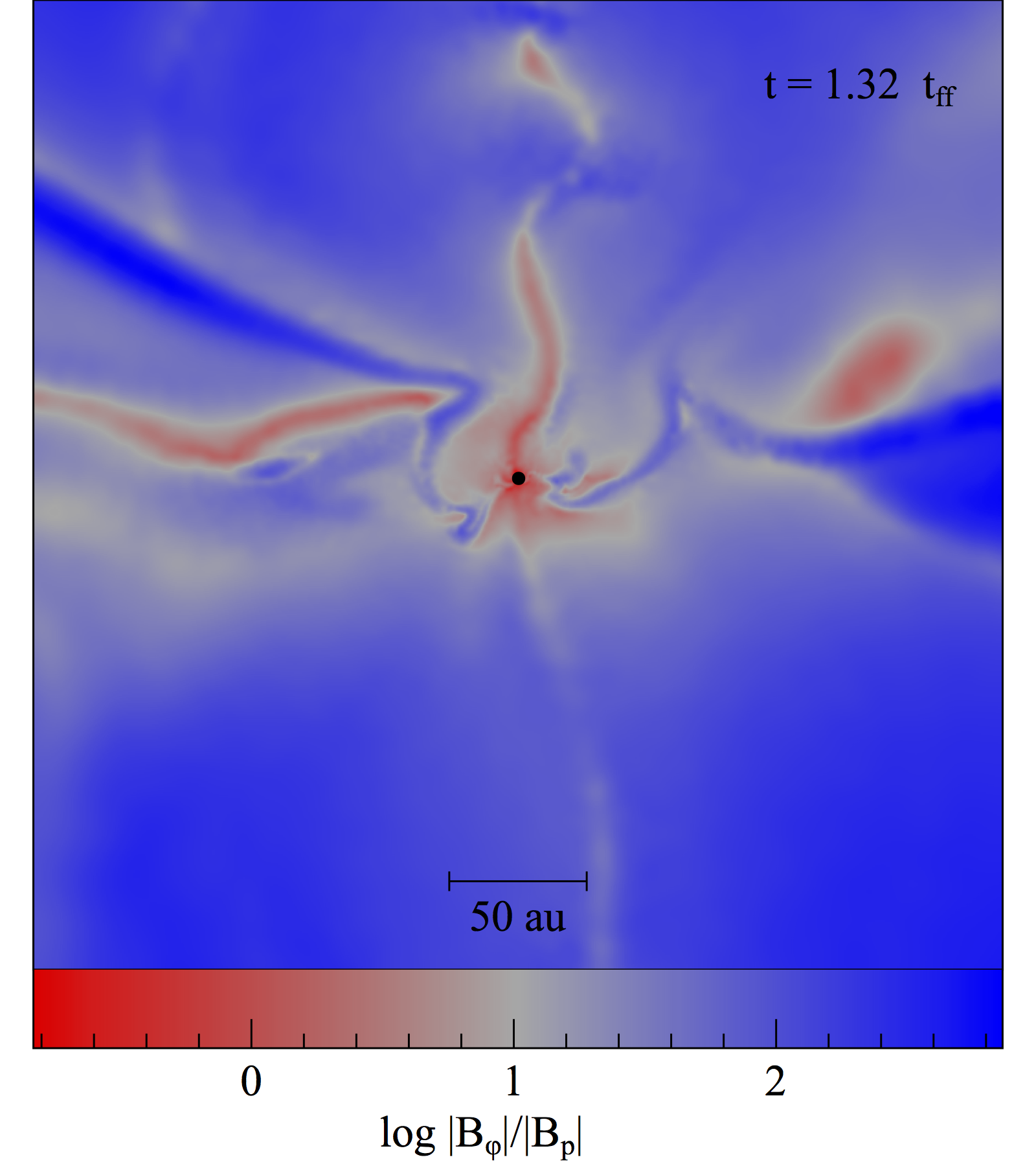}

\caption{Tranverse sections, in a $z$-$x$ plane through the sink particle of the ratio of the poloidal magnetic field component to the toroidal component for the three calculations presented in \cref{fig:stirred}. As in \cref{fig:shakenTP}, red indicates regions where the poloidal field component dominates compared to blue regions where the toroidal field is dominant.
\label{fig:stirredTP}}
\end{figure*}

We now decompose the magnetic field into cylindrical components\footnotemark{}, where
\begin{empheq}[left={\empheqlbrace}]{equation}\label{eqn:cylindricalcoords}
   \begin{aligned}
      r &= \sqrt{x^2 + y^2} \\
      \phi &= \arctan\thinspace\frac{y}{x}\\
      z &= z\comma
   \end{aligned}
\end{empheq}
so that the magnetic field in this co--ordinate system is given by
\begin{empheq}[left={\empheqlbrace}]{equation}\label{eqn:Bcylindricalcoords}
   \begin{aligned}
      B_r &= B_x \frac{x}{r} + B_y \frac{y}{r} \\
      B_\phi &= B_y \frac{x}{r} - B_x \frac{y}{r}\\
      B_z &= B_z\fstop
   \end{aligned}
\end{empheq} 
Then \citep[\eg, as in][]{1955ApJ...122..293P} the azimuthal field, $B_\phi$, and the toroidal field component are identical,
\begin{empheq}{equation}\label{eqn:torB}
   B_\rmn{tor}=B_\phi\comma
\end{empheq}
and the magitude of the poloidal component is given by
\begin{empheq}{equation}\label{eqn:polB}
   \left|B_\rmn{p}\right|= \sqrt{B^2_r + B^2_z}\fstop
\end{empheq}

\footnotetext{The form of this decomposition shown in equations (16--18) in\citet{2017MNRAS.467.3324L} was incorrect. However, the analysis and all plots used the correct equations as presented here.}

In \mylb we concluded that the essential ingredients for collimated outflows were a strong poloidal field to lift material out of the disc-plane \textit{coupled with} a strongly toroidal component to collimate the outflow. A solely torodial field trapped fluid in the disc and a solely poloidal field would produce a bulk outflow (\cf the \enquote{detachment} seen for a $\mu = 10$ calculation in \mylb when the poloidal component overhauls the torodial one). In \cref{fig:shakenTP} we show how this effect is mirrored for these calculations. Firstly, the sub-sonic $\mu = 20$ calculations have a strongly toroidal disc, with no significant region where the poloidal field dominates. 

Conversely, the $\mu = 5$ and $10$ calculations \textit{with} an outflow have a poloidally dominated region which is then wound-up by the strongly toroidal disc (exemplified by the toroidal inner region of the jet). Clearly, then, by disrupted the formation of a pseudo-disc, turbulence is able to prevent angular momentum transport via a bipolar outflow. However, unlike for a laminar core, this does not lead to fragmentation. Whilst fragmentation can be suppressed in stronger fields ($\mu \geq 10$) by magnetic braking, hitherto we found that all $\mu = 20$ discs were gravitationally unstable and liable to fragment into binary or ternary systems. We note, however, that the increased turbulent velocity field has the effect of both making the (pseudo--)discs less massive and also placing less angular momentum in the disc. Both of these effects make the disc less vulnerable to fragmentation from gravitational--rotational torques. This latter point is echoed by comparing the total rotational energy of the whole core, which is reduced in the transonic cores compared to the sub-sonic ones as the turbulent motion acts to prevent simple rotation around the $z$--axis. We note also that the \enquote{toroidal} region is a good tracer for the location of a pseudo--disc, albeit not the extent since this region will extend out the the edge of the core. For example, the transonic $\Mach = 1.0$ cores all develop discs which are neither flat nor close to perpendicular to the rotation axis (or equivalently perpendicular to the initial magnetic field axis).

~

The $\mu = 20$ calculations discussed in the preceding paragraph differ from the calculations presented in \citet{2017MNRAS.467.3324L} in an important way, viz. that here we include radiative transfer. This alone can prevent fragmentation of the core. However, to test whether the addition of turbulence alone can stop the fragmentation mode found in our previous work we also ran the $\mu = 20$ $\Mach = 1$ calculation \textit{without} the radiative transfer treatment (this is calculation $\upmu$20-M10(M)). We show the results of this calculation alongside $\upmu$20-M00(R) (i.e. a laminar, radiative, $\mu = 20$ calculation) in \cref{fig:mu20_Fragments}. Here we find that adding \textit{either} radiation or turbulence is sufficent to stabilise the core against fragmentation into a multiple system. This indicates that the fragmentation mode proposed in \citet{2017MNRAS.467.3324L} can be suppressed by making a warmer disc or by a smaller disc (or both).

From this, we conclude that highly turbulent initial conditions are --- all other things being equal --- inimical to the formation of bipolar jets and well defined pseudo-discs. This has consequences for observations: an observation that a very young protostellar object has a jet--like outflow implies that the core was initially sub--sonic. 

\section{Stirred: varying the initial angular momentum}
\label{sec:stir}

In \cref{sec:shaken} above, we showed that increasing the initial turbulent velocity field of the core into a transonic regime inhibits the formation of a (pseudo--)disc and therefore supresses the formation of a bipolar outflow. Since disc structures are inherently connected with the angular momentum (or somewhat equivalently the rotational kinetic energy) in a collapsing core, we now take the $\mu = 5$ calculations with a $\Mach =  1.0$ velocity field and increase the initial rotation rate. These are calculations $\upmu$05--M10--r001(R) and $\upmu$05--M10--r002(R) in \cref{tbl:initialcond}. $\upmu$05-M10(R) has $\beta_\rmn{rot} = 0.005$, so these two calculations have double and quadruple the initial rotational energy (and hence the angular momenta are $\sqrt{2}$ and double the initial values) respectively.

\Cref{fig:stirred} compares these three calculations at a similar evolutionary epoch. We find that $\upmu$05--M10(R)--r001 (the calculation with double the initial angular momentum) has a very weak outflow, corresponding to a confused disc--like structure; however, $\upmu$05--M10(R)--r002 produces a clear disc and a corresponding bipolar outflow. \Cref{fig:stirredvz} clearly shows this, with a region of fluid moving the outward from the sink particle, \ie where $v_z > 0$ for $z>0$ (above the plane of the disc) and vice versa. We note that material undergoing a gravitational self collapse invariably has $v_z \leq 0$ above the disc and vice versa. Unlike the sub--sonic cores presented above and the laminar cores presented in \mylb, this outflow is much slower (\ca $\left|v_z\right| \approx 1~\kvelo$) and much less substantial. The latter effect is shown by the reduced column density contrast between the outflow material and the surrounding cloud core.

The nature of any disc structure is related to the initial rotation rate. At the lowest initial rotation rate, i.e. calculation $\upmu$05--M10(R), no disc forms. Conversely at the highest rotation rate --- calculation $\upmu$05--M10--r002 --- a clear pseudo--disc has formed. However, this disc has a rotation speed slower than that seen in \cref{fig:radvel} for a sub--sonic core (and which is therefore even slower than a laminar core, e.g. $\upmu$05--M00).
This is the cause of the weaker outflow seen here. Nonetheless, this illustrates that whilst turbulent kinetic energy can disrupt a core so much that it does not form a pseudo--disc, adding additional \textit{rotational} energy can act to re--form that disc structure. To quantify this effect, we define a new parameter,
\begin{equation}
\epsilon = \frac{\beta_\rmn{turb}}{\beta_\rmn{rot}} \equiv \frac{E_\rmn{turb}}{E_\rmn{rot}}\comma
\end{equation}
such that $\epsilon > 1$ implies turbulence is the dominant source of kinetic energy, conversely $\epsilon < 1$ implies rotation is, and $\epsilon \approx 1$ implies approximate equality between the two fields. A value of $\epsilon \lesssim 1$ is a necessary pre-requisite for the formation of a pseudo--disc and outflow. The $\upmu$05-M10(R) calculation has $\epsilon = 26 \gg 1$ and correspondingly the pseudo--disc was highly disrupted; conversely $\upmu$05--M10--r002(R) had $\epsilon = 1.06 \approx 1$ and has formed a disc with a weak outflow. The intermediary phase $\upmu$05--10--r001(R) has $\epsilon = 1.6$ which corresponds, as expected, to a disrupted pseudo-disc. These values are compatible with those obtained for lower Mach numbers, \eg $\upmu$05--M01 and $\upmu$05--M03 have $\epsilon = 0.26 < 1$ and $\epsilon = 2.4 \approx 1$ respectively (identical values are necessarily obtained for 10-01, \etc). 

Comparing the outflow velocities between the earlier laminar cores, the sub--sonic cores in \cref{sec:shaken} and the transonic cores in this section we find a steady decrease in the outflow speed, $\left|v_z\right|$, as the conditions become more complicated. For example, a laminar core with a fully aligned magnetic field can produce an outflow with $\left|v_z\right| = 8~\kvelo$, a sub--sonic core where $\epsilon = 2.4$ produces a slower and less collimated outflow, but the transonic core with $\epsilon = 1.06$ only produces $\left|v_z\right| = 0.5~\kvelo$.

\Cref{fig:stirredTP} shows the same poloidal--toroidal decomposition as in 
\cref{fig:shakenTP} but for the three $\Mach = 1$ calculations discussed in this 
\namecref{sec:stir}. The change between no outflow when $\epsilon = 26$ and a collimated jet when $\epsilon = 1.06$ is again shown in this analysis. There is also the hint of a weak outflow in the highly distrupted pseudo--disc when $\epsilon = 1.6$, however, the lack of a disc structure to collimate this suppresses a jet structure from being formed. This highlights how the formation of a disc is necessary to produce a jet--like outflow around a first hydrostatic core.

\section{Discussion}
\label{sec:discussion}

We observe that the ratio of the total magnetic, turbulent and rotational kinetic energies to the gravitational self-potential in the $\Mach =  1$ cores is very close to unity. Consequently, these cores are on the limit of boundedness and would become unbound with only a small increase in the turbulent velocities. 
This is consistent with the observed average velocities seen in cluster scale simulations, where cores (or comparable regions) undergoing a Jeans collapse are often sub--sonic, notwithstanding the initial highly super--sonic velocities in the progenitor cloud \citep*{2000ApJ...535..887K}, and compatible with observed star formation regions, e.g. in Ophiuchus \citet{2015MNRAS.450.1094P}.

A large value of $\epsilon$ in these highly turbulent (and weakly bound) cores is comparable to the strongly misaligned magnetic field and rotation axes that we explored in \mylb. In those cases, values of the parameter $\vartheta \geq 60\degree$ produced highly disrupted discs --- with correspondingly slower rotation profiles --- and this was fatal to the production of an outflow. The production of a bipolar jet from a first hydrostatic core is therefore linked to both having a suitable field strength and geometry (turbulence can not make a weaker field more able to produce the necessary poloidal components to form a jet) to generate the outflow at all, combined with an environment conducive to the formation of a sub-Keplerian pseudo--disc. This former requirement is satisfied by having \hl{$\mu \leq 10$} and $\vartheta < 60\degree$, whilst the latter requires $\epsilon \lesssim 1$ which implies either a sufficently weak turbulent velocity field \textit{or} a sufficiently strong rotational velocity field.

Observations of potential first hydrostatic core outflows \citep[\eg][]{2011ApJ...742....1D} report outflow speeds of around $\left|v_z\right| = 5~\kvelo$. These are consistent with the observed $\left|v_z\right|$ values seen for very weakly turbulent and laminar cores, where $\epsilon \ll 1$ but in these configurations the total kinetic energy is actually relatively low. Outflows are produced from much higher initial kinetic energies, i.e. when $\Mach = 1$, provided that $\epsilon \lesssim 1$. However, the outflow produced in this scenario --- transonic turbulence coupled with additional angular momentum --- is much slower. This has an important implication for observations, \viz that faster outflows are the result of less `dynamic' initial conditions. 
Conversely it also indicates that a slower outflow speed is a result of an elevated value of $\epsilon$.
This allows a limited degree of re--winding of the evolution of very young protostellar objects to predict the conditions in the Jeans unstable core from which they formed. The mass and size of the first hydrostatic core provides a limit on the maximum outflow and jet velocity (see \citealp{2003MNRAS.339.1223P}). Therefore, simply trying to decrease $\epsilon$ to obtain faster jets is not possible, and objects such as MMS-6/OMC-3 found by \citet{2012ApJ...745L..10T} or the Herbig--Haro object HH 1165 found by \citet{2017arXiv170501170R} are more evolved than the first core phase. 

We noted earlier that our results are broadly comparable with \citet{2017arXiv170309139M}. For example, in the initial phase of the collapse we obtain a similar link between oblateness and field strength, and disruption of the collapsing core by higher Mach numbers \citep[see also][]{2011ApJ...728...47M}. However, \citet{2017arXiv170309139M} use significantly stronger magnetic fields ($1.12 \leq \mu \leq 2.81$). The core mass used there was also higher than our work ($M_\rmn{core} = 2.51~\solarm$ as opposed to $1~\solarm$) so this implies a much stronger magnetic field is present. We obtain a similar reduction in the radius of the pseudo--disc as $\mu$ is reduced as in that paper, however, because we also probe much weaker field strengths --- \hl{$\mu \geq 20$} --- we also find that once magnetic effects are reduced further the disc size again decreases due to reduced angular momentum transport \citep{2017MNRAS.467.3324L}. The difference in magnetic field strengths also explains the lack of any magnetic pressure induced \enquote{cavity} in our calculations. We also note that our calculations include a radiative transfer scheme and begin with uniform density, not Bonnor--Ebert, spheres. 

We find a strong, but not perfect, alignment between our pseudo--discs and the outflow produced. This is likely due to the way we super--impose a random turbulent velocity and a solid body rotation profile together. This guarantees that the angular momentum vector of the system will be very closely aligned to the $z$--axis (which is also the magnetic field axis). In comparison, an approach whereby the random seed used to generate the turbulence is used to produce angular momentum will result in a more complicated angular momentum vector. Consequently, in such a system misalignements between the discs and outflows are more likely to occur. However, our approach has the advantage that it is possible to separate linear and rotational effects more easily. Keeping the rotational component of the velocity field separate would also allow the effect of varying $\vartheta$ to be more readily explored. We note, however, that the partially asymmetrical appearance of the outflow when $\Mach = 1$ and $\epsilon \gg 1$ is compatible with the complex structure of HH 1165 \citep{2017arXiv170501170R}, and in particular the asymmetrical nature of the outflow seen in that system. This indicates that the combination of rotational effects --- to create a disc and outflows --- and turbulence is necessary to explain the complex structures seen in protostars.  

\section{Conclusions}
\label{sec:conclusions}

We performed eighteen smoothed particle magnetohydrodynamical calculations of the collapse of a molecular cloud core. Four of these calculations used a barotropic equation of state; the remaining fourteen use a flux--limited diffusion radiative transfer scheme. These calculations use mass--to--flux ratios ranging from $5 \leq \mu \leq 20$, initial turbulent Mach numbers $0 \leq \Mach \leq 1$ and initial rotation rates corresponding to $0.005 \leq \beta_\rmn{rot} \leq 0.02$.  

Calculations using radiative transfer produces a smoother distribution of thermal energy and hence promote the formation of larger, warmer, discs compared to the barotropic equation of state. Consequently the use of a full RMHD scheme is desireable in calculations of protostellar collapses.

We obtain a strong link between the degree of turbulence in the initial molecular core and the nature of the consequent outflow. Laminar, $\Mach = 0$, cores (now with radiative transfer) produce results comparable to our earlier work \citep[see][]{2017MNRAS.467.3324L} with a strong $\left|v_z\right| \simeq 5\thinspace\kvelo$ bipolar jet and outflow for strong magnetic fields ($\mu \leq 10$). No collimated or substantial outflow is produced for $\Mach = 0$, $\mu = 20$.  At the other extreme, when $\beta_\rmn{rot} = 0.005$ as before but $\Mach = 1$ the calculation is highly distrupted and no outflow is produced for all field strengths. The addition of turbulence or radiative transfer, or both, can also stabilise a $\mu = 20$ core against fragmentation into binary or ternary systems.

We then increase the initial rotation rate from $\beta_\rmn{rot} = 0.005$ to $\beta_\rmn{rot} = 0.02$ for $\Mach = 1$. The additional angular momentum promotes the formation of a disc when the ratio of turbulent kinetic to rotational kinetic energy, $\epsilon \lesssim 1$. Once such a disc is formed, we again find that an outflow is produced albeit with a slower velocity of $\left|v_z\right| \simeq 1\thinspace\kvelo$ and with some asymmetries. We therefore conclude that the velocity of a first hydrostatic core outflow is related to this ratio, $\epsilon$, and that faster outflows are correlated with lower values of $\epsilon$ and vice versa.

\section*{Acknowledgments}

We thank the anonymous referee for their comments and suggestions which have resulted in an improved manuscript.

BTL acknowledges support from an STFC Studentship. 
MRB was supported by the European Research Council under the European Commission's Seventh Framework Programme (FP7/2007-2013 Grant Agreement No.
339248). 

This work used the DiRAC Complexity system, operated by the University of Leicester IT Services, which forms part of the STFC DiRAC HPC Facility (\url{www.dirac.ac.uk}). This equipment is funded by BIS National E-Infrastructure capital grant ST/K000373/1 and STFC DiRAC Operations grant ST/K0003259/1. DiRAC is part of the National E-Infrastructure.

Calculations were also performed on the University of Exeter Supercomputer, a DiRAC Facility jointly funded by STFC, the Large Facilities Capital Fund of BIS and the University of Exeter.

This work also made use of the \textsc{numpy} \citep*{walt2011numpy} and \textsc{matplotlib} \citep{hunter2007matplotlib} \textsc{python} modules. Rendered figures were produced using the \textsc{splash} \citep{2007PASA...24..159P,2011ascl.soft03004P} visualisation programme. 

\renewcommand*{\refname}{REFERENCES}
\bibliographystyle{mnras}
\bibliography{Turbulence}

\bsp

\label{lastpage}

\end{document}